\documentclass[english,aps,pre,showpacs,superscriptaddress,twocolumn,floats,amssymb]{revtex4}
\usepackage{amsmath}
\usepackage{graphicx}

\makeatletter
\@ifundefined{textcolor}{}
{%
 \definecolor{BLACK}{gray}{0}
 \definecolor{WHITE}{gray}{1}
 \definecolor{RED}{rgb}{1,0,0}
 \definecolor{GREEN}{rgb}{0,1,0}
 \definecolor{BLUE}{rgb}{0,0,1}
 \definecolor{CYAN}{cmyk}{1,0,0,0}
 \definecolor{MAGENTA}{cmyk}{0,1,0,0}
 \definecolor{YELLOW}{cmyk}{0,0,1,0}
 }

\usepackage{graphicx}
\usepackage{amsmath}
\usepackage{amssymb}
\bibstyle{apsrev.bib}

\newcommand{\beq}{\begin{equation}}
\newcommand{\eeq}{\end{equation}}

\newcommand{\comment}[1]{}

\usepackage{babel}

\usepackage{babel}

\makeatother

\usepackage{babel}

\begin{document}

\title{Ground States of the Sherrington-Kirkpatrick Spin Glass with Levy
Bonds}
\author{Stefan Boettcher}
\email{sboettc@emory.edu}
\affiliation{Physics Department, Emory University, Atlanta, Georgia 30322, USA}
\begin{abstract}
Ground states of Ising spin glasses on fully connected graphs are
studied for a broadly distributed bond family. In particular, bonds
$J$ distributed according to a Levy distribution $P(J)\propto1/|J|^{1+\alpha},$
$|J|>1,$ are investigated for a range of powers $\alpha.$ The results
are compared with those for the Sherrington-Kirkpatrick (SK) model,
where bonds are Gaussian distributed. In particular, we determine
the variation of the ground state energy densities with $\alpha$,
their finite-size corrections, measure their fluctuations, and analyze
the local field distribution. We find that the energies themselves
at infinite system size attain universally the Parisi-energy of the
SK as long as the second moment of $P\left(J\right)$ exists ($\alpha>2$),
and compare favorably with recent one-step replica symmetry breaking
predictions well below $\alpha=2$. At and just below $\alpha=2$,
the simulations deviate significantly from theoretical expectations.
The finite-size investigation reveals that the corrections exponent
$\omega$ decays from the SK value $\omega_{SK}=\frac{2}{3}$ already
well above $\alpha=2$, at which point it reaches a minimum. This
result is justified with a speculative calculation of a random energy
model with Levy bonds. The exponent $\rho$ that describes the variations
of the ground state energy fluctuations with system size decays monotonically
from its SK value over the entire range of $\alpha$ and apparently
vanishes at $\alpha=1$. 
\end{abstract}


\maketitle

\section{Introduction\label{sec:Introduction}}

The mean-field spin glass, in particular, the Sherrington-Kirkpatrick
model (SK)~\citep{Sherrington75} provides the most thorough and
tantalizing insights into the nature of frustrated and disordered
systems~\citep{MPV,F+H}. The applications of this model, and the
techniques that have been developed to dissect it, have dramatically
expanded since its inception, and pervade many areas of science ~\citep{MPV,Nishimori01,MPZ,Talagrand03,Eastham06,Schneidman06,Percus06,Mezard06}.
Yet, even for the SK model, many essential aspects still have to be
revealed. In particular, finite size corrections to the mean-field
solutions have not been worked out, and existing conjectures appear
to be inconsistent or disagree with numerical predictions
~\citep{Bouchaud03,Aspelmeier03,Andreanov04,EOSK,Aspelmeier07,Palassini08,Boettcher10b,Boettcher10a}.
These questions are not merely academic but seem to indicate an incomplete
understanding in how finite-dimensional spin glasses, as introduced
by Edwards and Anderson~\citep{Edwards75}, approach the large-dimensional
limit~\citep{Aspelmeier03,Boettcher05d}.

The study of spin glasses with a power-law (Levy) bond distribution
has been advocated in Ref.~\citep{Cizeau93} on physical grounds,
concerning the properties of the RKKY couplings between magnetic sites
in the dilute limit of glassy alloys. In fact, it thus may constitute
a more {}``physical'' mean-field limit compared to the typical,
mathematically more tractable, Gaussian bonds used in SK. One could
speculate to whether power-law bonds may actually overcome the discrepancies
between finite-dimensional spin glasses and their mean-field limit.
While power-law bonds do not seem to affect the phenomenology of the
droplet model proposed for low-dimensional lattices, the corresponding
(replica symmetric) mean-field theory in Ref.~\citep{Cizeau93} shows
much weaker replica symmetry breaking effects at low temperatures
than SK, which is more in line with many numerical and experimental
results in finite dimensions. An in-depth study of the low-temperature
properties is pertinent as Levy spin glasses have received much renewed
theoretical attention recently\citep{Janzen08,Janzen10a,Janzen10b,Neri10}.
In particular, Ref.~\citep{Janzen10b} provides a number of theoretical
predictions for the ground state energy density both, at the replica
symmetric (RS) and the one-step replica symmetry breaking (1RSB) level,
to compare with. 

In this paper, we probe mean-field spin glasses for a one-parameter
family of bond distributions with power-law tails to explore how this
may affect familiar properties, such as the distribution of ground
state energies, its characteristic width, and the finite-size scaling
corrections to the energy in the thermodynamic limit. In particular,
we consider bonds $J$ distributed according to 
\begin{eqnarray}
P(J) & = & \frac{\alpha}{2}\left|J\right|^{-1-\alpha},\qquad|J|>1,
\label{eq:powerlaw}
\end{eqnarray}
for a range of powers $\alpha.$ (In contrast to most theoretical
work that is limited to specific regimes of $\alpha$, we do not rescale
the bonds with system size $N$ in our simulations; proper densities
for the various regimes are defined in Sec.~\ref{sec:PowerLaw-Spin-Glasses}.)
From a conceptual standpoint, the bond distribution in Eq.~(\ref{eq:powerlaw})
is interesting because it provides a gradual interpolation between
those distributions possessing a second moment (for $\alpha>2)$ and
those that don't, i.e., where that moment diverges with the system
size $N.$ For instance, singular behavior for the ground state energy
has been obtained near $\alpha=1$ and 2 in Ref.~\citep{Janzen10b},
at least at the replica symmetric level. While we expect the behavior
for $\alpha>2$ to be similar to SK, it is interesting to see how
this limit is approached and how scaling exponents may change with
$\alpha$. The extreme limit of bonds well-separated in size have
been studied in Refs.~\citep{Newman94,Cieplak94}.

In our simulations, we use \emph{extremal optimization} (EO)~\citep{Boettcher01a,Dagstuhl04},
a local search heuristic which has been used successfully to obtain
ground state approximations for mean-field and finite-dimensional
spin glasses with bimodal and Gaussian bond distributions~\citep{EOSK,Boettcher10a,Boettcher10b}.
Broadly distributed bonds create a very heterogeneous configuration
space, which provides a significant challenge for any local search
heuristic, and considerable computational resources had to be expanded
to obtain sufficiently accurate results. Combined with the desire
to sample an entire $\alpha$-family of models, such obstacles have
limited the achievable system sizes, which may at times call into
question whether true asymptotic behavior has been reached. At their
face value, the EO results are consistent with RSB predictions for
the ground state energies except near $\alpha=2$, where EO appears
to provide a finite value. Both, the exponent for finite-size corrections
$\omega$ and for the width of energy fluctuations $\rho$, exhibits
an interesting variation with $\alpha$ at $T=0$, most dramatically
near $\alpha=2$. In fact, due to their higher-order nature, these
variations in finite-size effects are already noticeable below $\alpha=4$,
well before the disappearance of the second moment in the bond distribution.
That such finite-size effects may exhibit non-universal (i.e., bond
distribution dependent) behavior has already been observed for mean-field
spin glasses on sparse graphs~\citep{Boettcher10a,Zdeborova10}. 

This paper is organized as follows: In the next section, we describe
the model and our conventions for the determination of the energy
densities. In Sec.~\ref{sec:Finding-Ground-States}, we briefly describe
the EO-heuristic used to obtain the numerical results, which are subsequently
discussed in detail in Sec.~\ref{sec:numerics}. In the Appendix,
we outline our speculative (and somewhat lengthy) calculation for
the random energy model (REM)~\citep{derrida:80,derrida:81}, which
we use in the discussion of the numerical results.

\section{Spin Glasses with Power-Law Bonds\label{sec:PowerLaw-Spin-Glasses}}

The Hamiltonian for a fully connected mean-field spin glasses is defined
as \begin{equation}
H=\frac{1}{{\cal N}}\sum_{i=1}^{N-1}\sum_{j=i+1}^{N}J_{i,j}\sigma_{i}\sigma_{j},\label{eq:Hamiltonian}\end{equation}
where we formally included a normalization factor ${\cal N}$ to make
the energy extensive, in light of the fact that the bonds $J_{i,j}$
are taken from the unrescaled density in Eq.~(\ref{eq:powerlaw}).
As we are here exclusively interested in properties of the ground
state energy density, it is convenient to simply set ${\cal N}=1$
and accept the fact that the putative ground state energies $E$ found
in our simulations for instances with the Hamiltonian in Eq.~(\ref{eq:Hamiltonian})
are neither extensive nor properly adjusted to the characteristic
coupling strength $J_{0}$. We obtain proper energy densities, independent
of system size $N$, in the relevant regimes of $\alpha$ via 
\begin{equation}
e=\begin{cases}
\frac{E}{N^{1+\frac{1}{\alpha}}}, & \alpha<2,\\
\frac{E}{N^{3/2}}, & \alpha>2.\end{cases}
\label{eq:energydensity}
\end{equation}
Note that in Eq.~(\ref{eq:energydensity}) we have chosen to ignore,
for all $\alpha>2$, the usual scaling with the second moment of $P(J)$,
i.e., instead of having a unit moment we have $J_{0}^{2}=\left\langle J^{2}\right\rangle =\frac{\alpha}{\alpha-2}$,
which fails to exist for $\alpha\leq2$. This choice allows us to
consider the special case $\alpha=2$ more closely, but to compare
with the familiar result from SK, $e_{SK}=-0.76321\ldots$
\citep{Oppermann05,Oppermann07}, we will have to take this extra
factor into account.

\section{Finding Ground States with Extremal Optimization\label{sec:Finding-Ground-States}}

In our numerical investigations here, we use the \emph{Extremal Optimization}
heuristic (EO)~\citep{Boettcher00,Boettcher01a}, in particular,
an adaptation of EO developed previously for SK~\citep{EOSK}.\footnote{The SK code is available in the DEMO section at http://www.physics.emory.edu/faculty/boettcher.} Due to the potentially wide distribution
of bonds (in particular, for small $\alpha)$, we pre-process the
bond matrix $J_{i,j}$ in a manner discussed in more detail in Ref.
\citep{Boettcher08a}: Any spin $\sigma_i$, $1\leq i\leq N$, is
\emph{determined}, subject to the state of spin $\sigma_{j'}$, if
\begin{equation}
\frac{1}{2}\sum_{j=1}^{N}\left|J_{i,j}\right|<{\textrm{max}}_{1\leq
  j\leq N}\left\{ \left|J_{i,j}\right|\right\} \equiv J_{i,j'},
\label{eq:dominantbond}
\end{equation}
such that the bond $J_{i,j'}$ is satisfied in the
ground state. This dependency allows us to eliminate the $i$th column
and row from the bond matrix by compounding the remaining bonds $J_{i,j\not=j'}$
according to 
\begin{eqnarray*}
J_{j',j} & \leftarrow &
J_{j',j}+{\textrm{sign}}\left(J_{i,j'}\right)J_{i,j}\qquad({\textrm{for~all~}}j\not=j')
\end{eqnarray*}
and similarly for $J_{j,j'},$ to preserve the symmetry of the bond
matrix. The ground state energy $E_{0}^{(N-1)}$ of this \emph{reduced}
bond matrix for $N-1$ spins gives the energy of the original system
via \[ E_{0}^{(N)}=E_{0}^{(N-1)}-\left|J_{i,j'}\right|.\]
Of course, this reduction procedure can be applied recursively until
there are no more spins $i$ that satisfies Eq.~(\ref{eq:dominantbond}).

Clearly, this reduction procedure is most effective for small $\alpha;$
for instance, at $\alpha=1/2,$ about half of all spins can be traced
out in this way. The resulting problem is not only drastically smaller
in size (considering the exponentially growing cost for determining
the ground state exactly!), but the bond matrix is also much more
\emph{homogeneous} for the subsequent application of the EO heuristic.
While the reduction is ineffective for $\alpha>2,$ even for, say,
$\alpha=3/2,$ where only a few spins get reduced, the elimination
of those exceedingly dominant bonds can be quite helpful for local
search with EO.

The remaining bond matrix is then treated by $\tau$-EO as outlined
in Ref.~\citep{EOSK}: A fitness proportional to the (negative of)
the local field is defined for each spin, and spins are chosen for
an update with a bias toward spins of low fitness as specified by
the parameter $\tau.$ For the system sizes in this paper, a value
of $\tau=1.5$ and a sequence of $O(N^{3})$ update steps for each
instance proved most effective. While systems with $\alpha<1$ are
readily reducible, and systems with $\alpha>2$ are very homogeneous,
those almost irreducible, but quite heterogeneous systems with $1<\alpha<2$
provide the biggest challenge for EO, requiring at least ten times
more update steps for consistent results. Each instance was treated
repeatedly with EO, each repeat starting from random initial conditions,
and results were considered consistent when a total of six runs reproduced
the putative ground state energy.

\section{Discussion of the Numerical Results\label{sec:numerics}}

We have determined approximate ground states for a large number of
fully connected graphs for sizes $N=31,\ldots,255$ with bond matrices
filled with random bonds drawn from the power-law distribution in
Eq.~(\ref{eq:powerlaw}) for values ranging over $\alpha=0.9,1.2,\ldots,3.0$
in steps of 0.3, and $\alpha=3.9$. Extremely large statistics was
required to obtain converging averages, which ultimately limited the
system sizes reached. For comparison, we have included earlier results
for SK~\citep{EOSK,Boettcher10b}, where sizes of $N=1023$ had been
obtained. At these finite system sizes, instances revealing the characteristics
of the thermodynamic limit are few and far between, as very large
bonds are an essential feature of the ensemble for $\alpha<2$ but
only arise infrequently. Hence, a large number of instances are required
for converging results. Here we averaged over about $10^{6}$ at the
smaller system sizes and at least $25,000$ for the largest sizes,
$N=255$. 

\begin{figure*}
\includegraphics[scale=0.33]{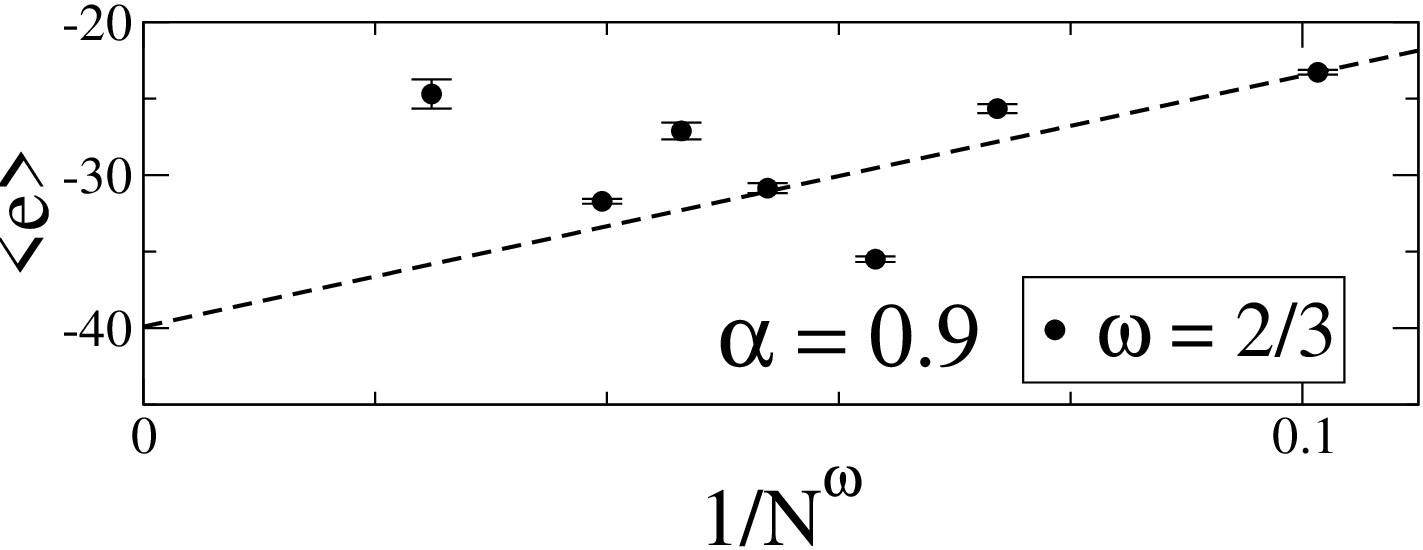}\hfill{}\includegraphics[scale=0.33]{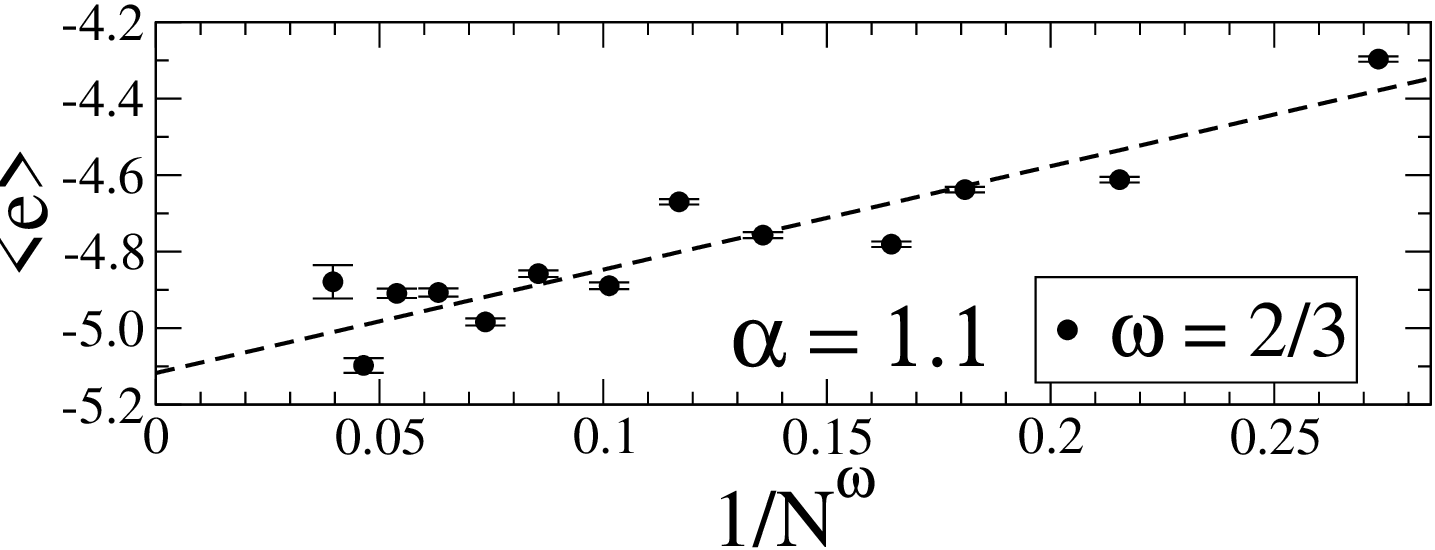}
\medskip{}
\includegraphics[scale=0.33]{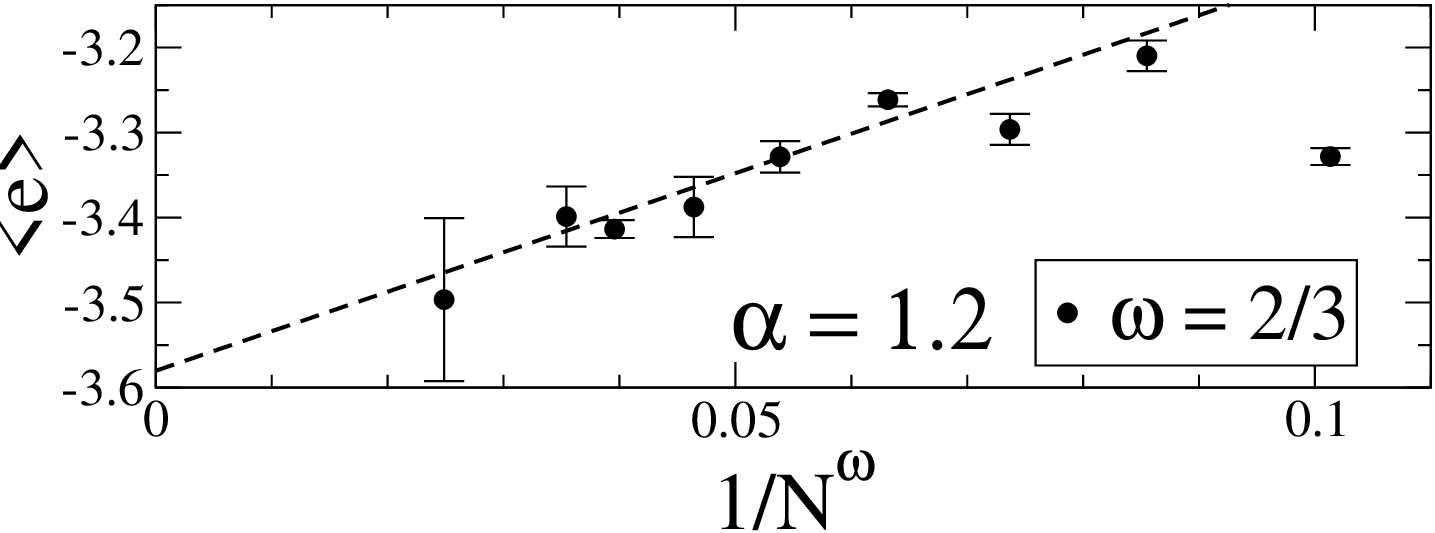}\hfill{}\includegraphics[scale=0.33]{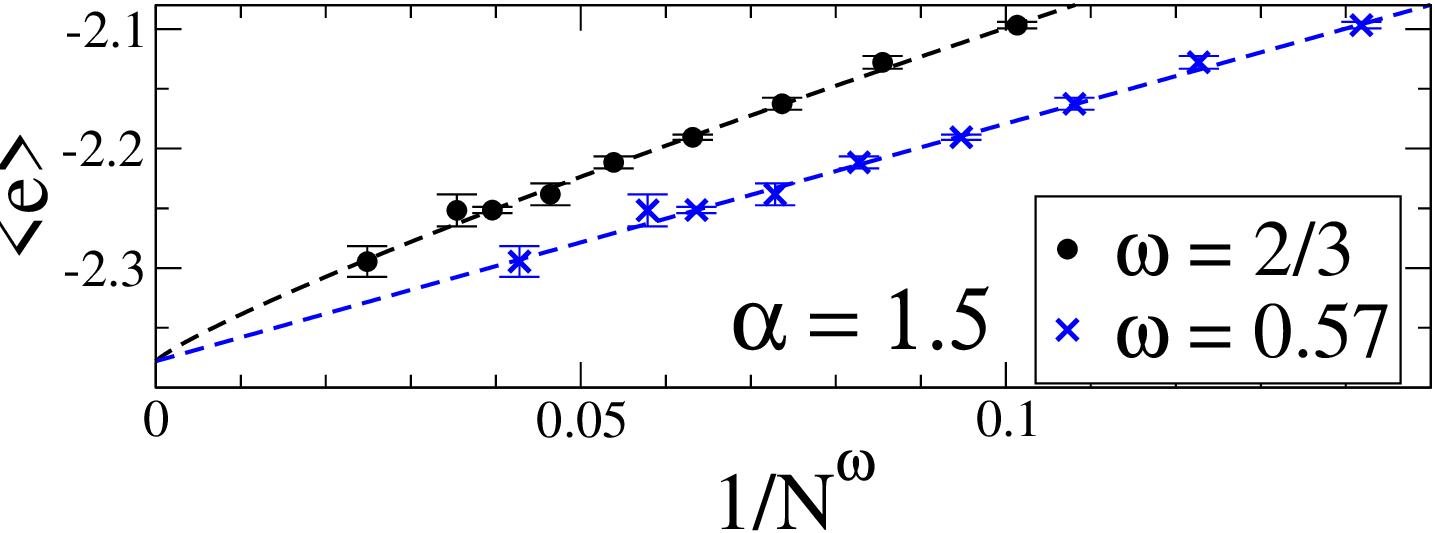}
\medskip{}
\includegraphics[scale=0.33]{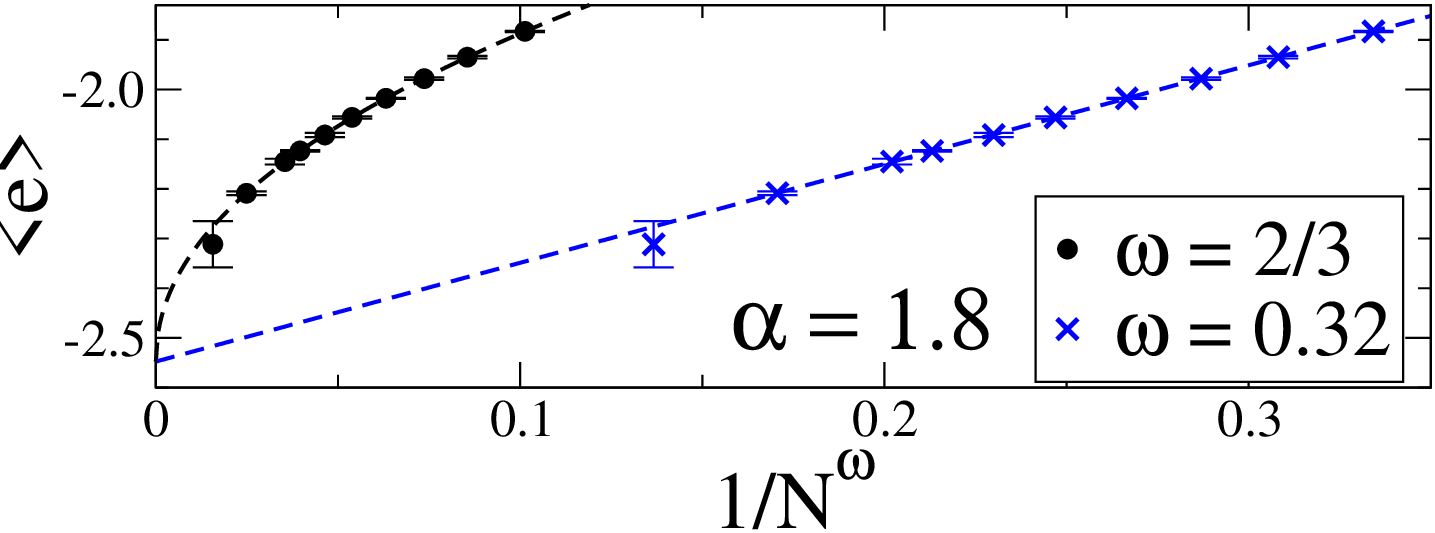}\hfill{}\includegraphics[scale=0.33]{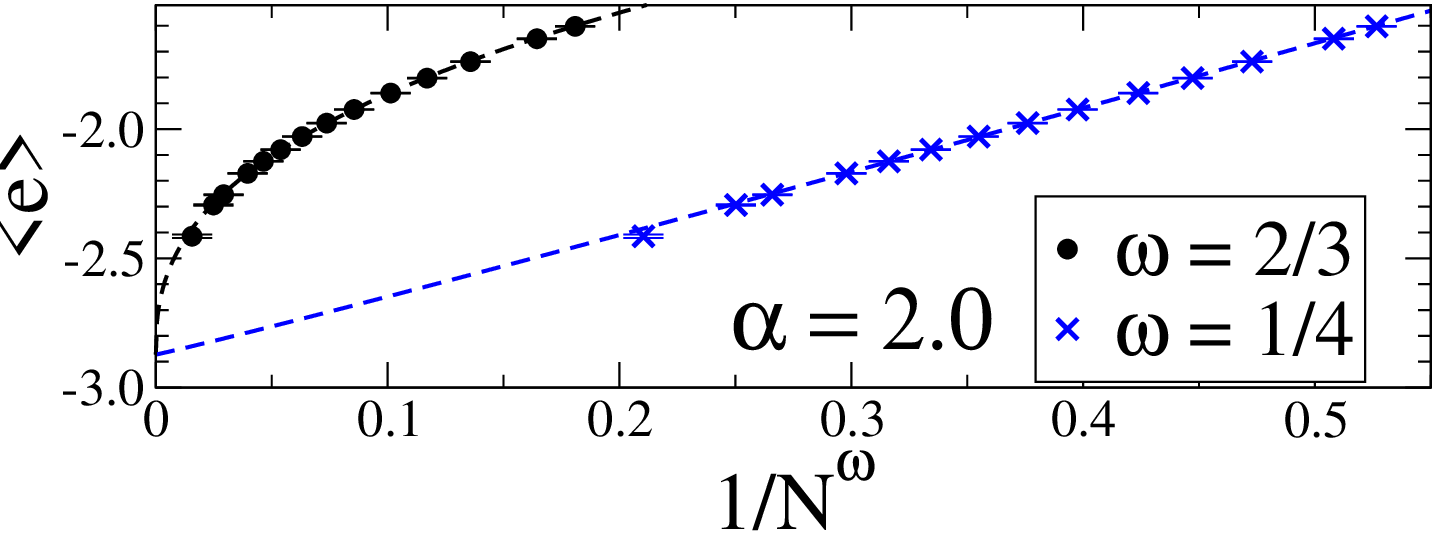}
\medskip{}
\includegraphics[scale=0.33]{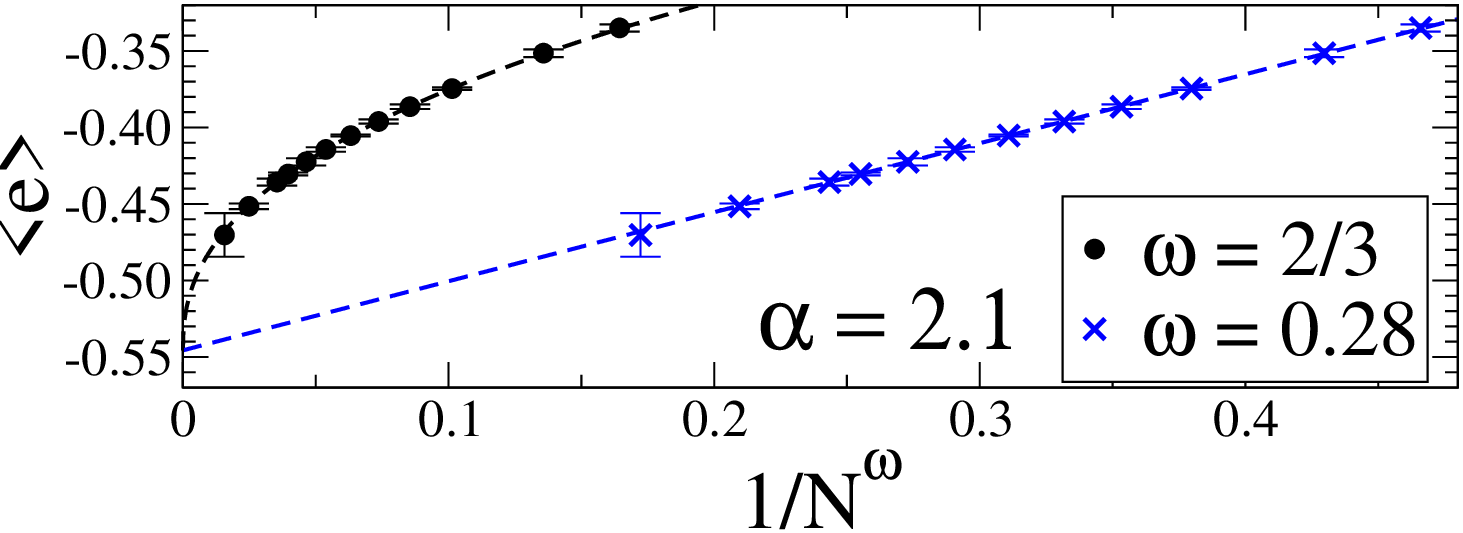}\hfill{}\includegraphics[scale=0.33]{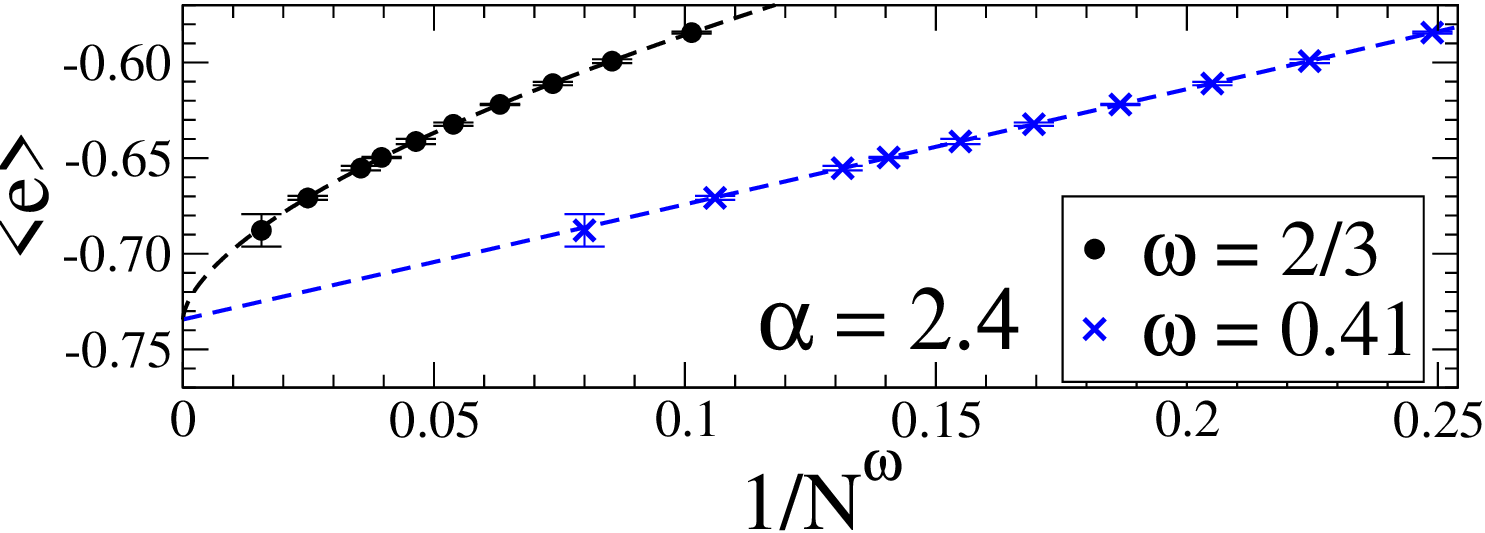}
\medskip{}
\includegraphics[scale=0.33]{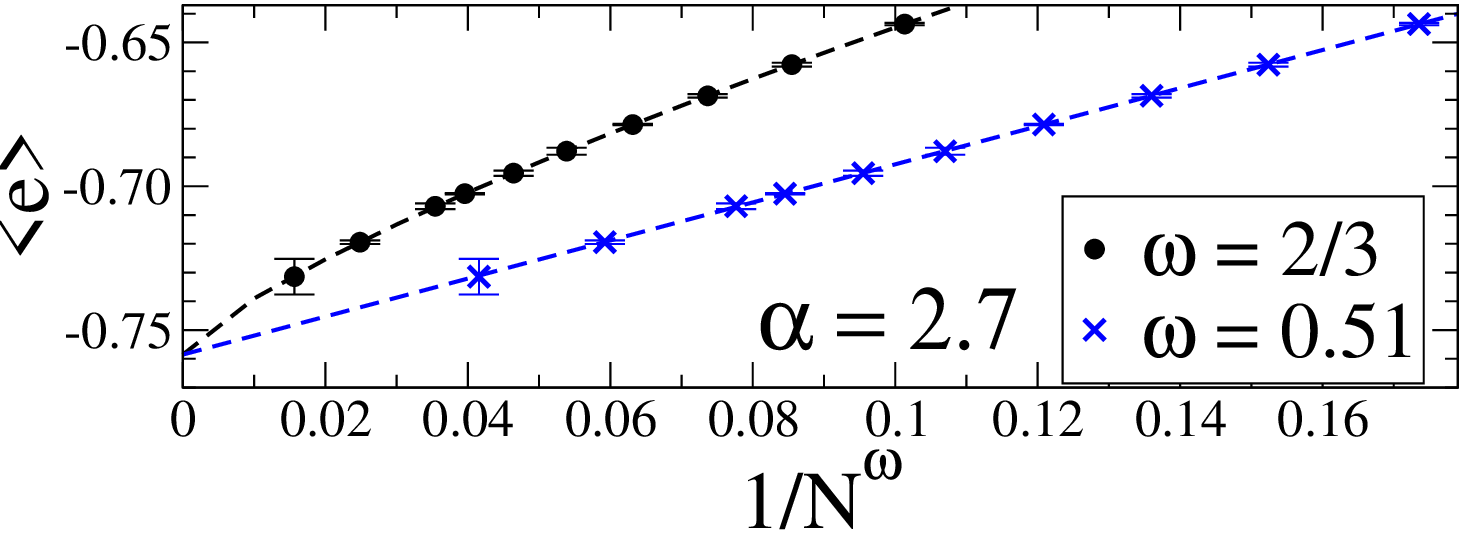}\hfill{}\includegraphics[scale=0.33]{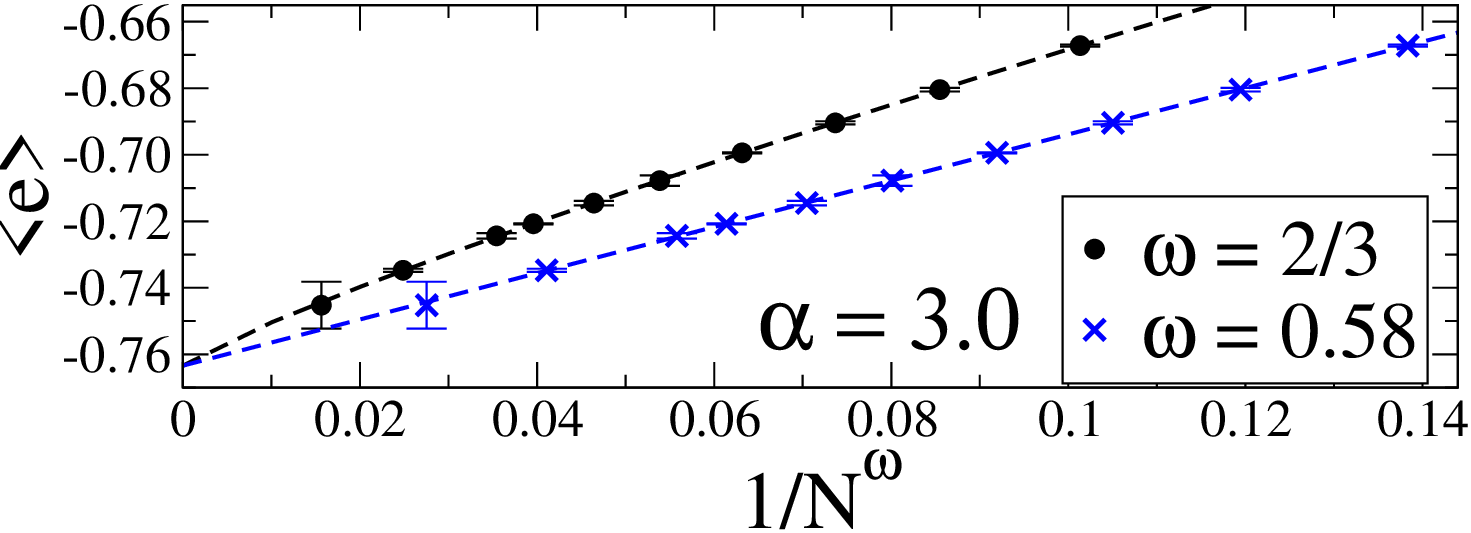}
\medskip{}
\includegraphics[scale=0.33]{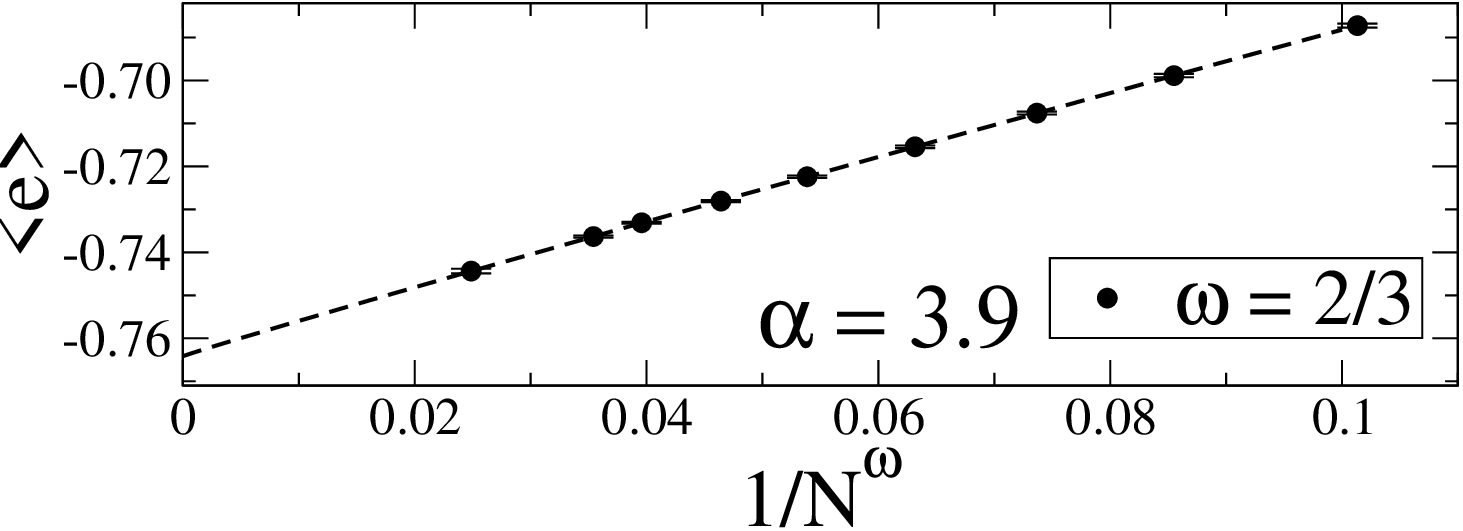}
\hspace*{\fill}
\caption{\label{fig:averageE}Average ground state energy density
  $\left\langle e\right\rangle $ obtained with EO for various system
  sizes up to $N=511$, for values of $\alpha$ between 0.9 and 3.9 in
  Eq.~(\ref{eq:powerlaw}). In each panel, the EO data is once plotted
  as a function of $1/N^{2/3}$ (black squares), the scaling projected
  for SK (corresponding to $\alpha\to\infty$), and once linearized
  with the fitted scaling correction exponent
  $\omega\left(\alpha\right)$ (blue $\times)$. While no specific
  scaling can be identified for the smallest $\alpha$, the scaling is
  consistent with SK (linear on this scale) for larger values of
  $\alpha$, but significant deviations from $\omega=2/3$ in
  Eq.~(\ref{eq:finitesize}) are observed for intermediate values of
  $\alpha$, where $\omega$ appears to attain a minimum of
  $\omega\approx0.25$ at $\alpha=2$. The results for
  $\omega\left(\alpha\right)$ are also summarized in
  Fig.~\ref{fig:omegarho}.}
\end{figure*}

\subsection{Corrections to scaling}

First, we investigate the average ground state energy density
$\left\langle e\right\rangle $ and its finite-size corrections to
scaling. In Figs.~\ref{fig:averageE} we plot these energies in an
extrapolation plot as a function of inverse system size, $1/N^{2/3}$,
which is generally believed to be the magnitude of scaling corrections
in the SK. We notice significant deviation from that scaling behavior
for varying $\alpha$. For each value of $\alpha$, we have fitted the
data to
\begin{eqnarray}
\left\langle e\right\rangle _{N} & \sim & \left\langle e\right\rangle
_{\infty}+\frac{A}{N^{\omega}}.
\label{eq:finitesize}
\end{eqnarray}
A plot of the same data but linearized through the scaling in Eq.
(\ref{eq:finitesize}) with the exponent $\omega$ extracted from
those fits are also shown in Fig.~\ref{fig:averageE}, and the fitted
values for $\omega$ as a function of $\alpha$ are displayed in Fig.
\ref{fig:omegarho}.

\begin{figure*}
\includegraphics[scale=0.25]{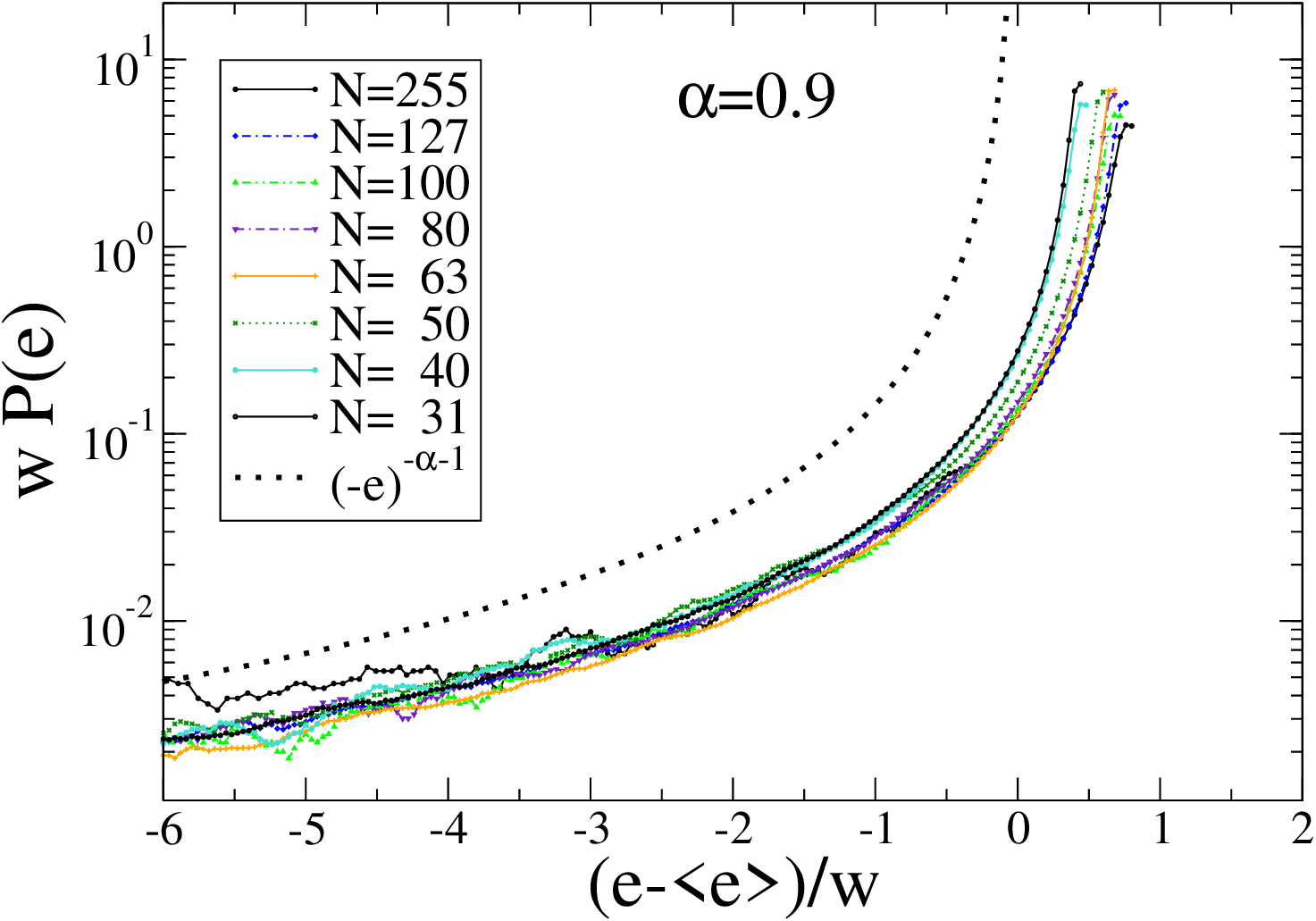}\includegraphics[scale=0.25]{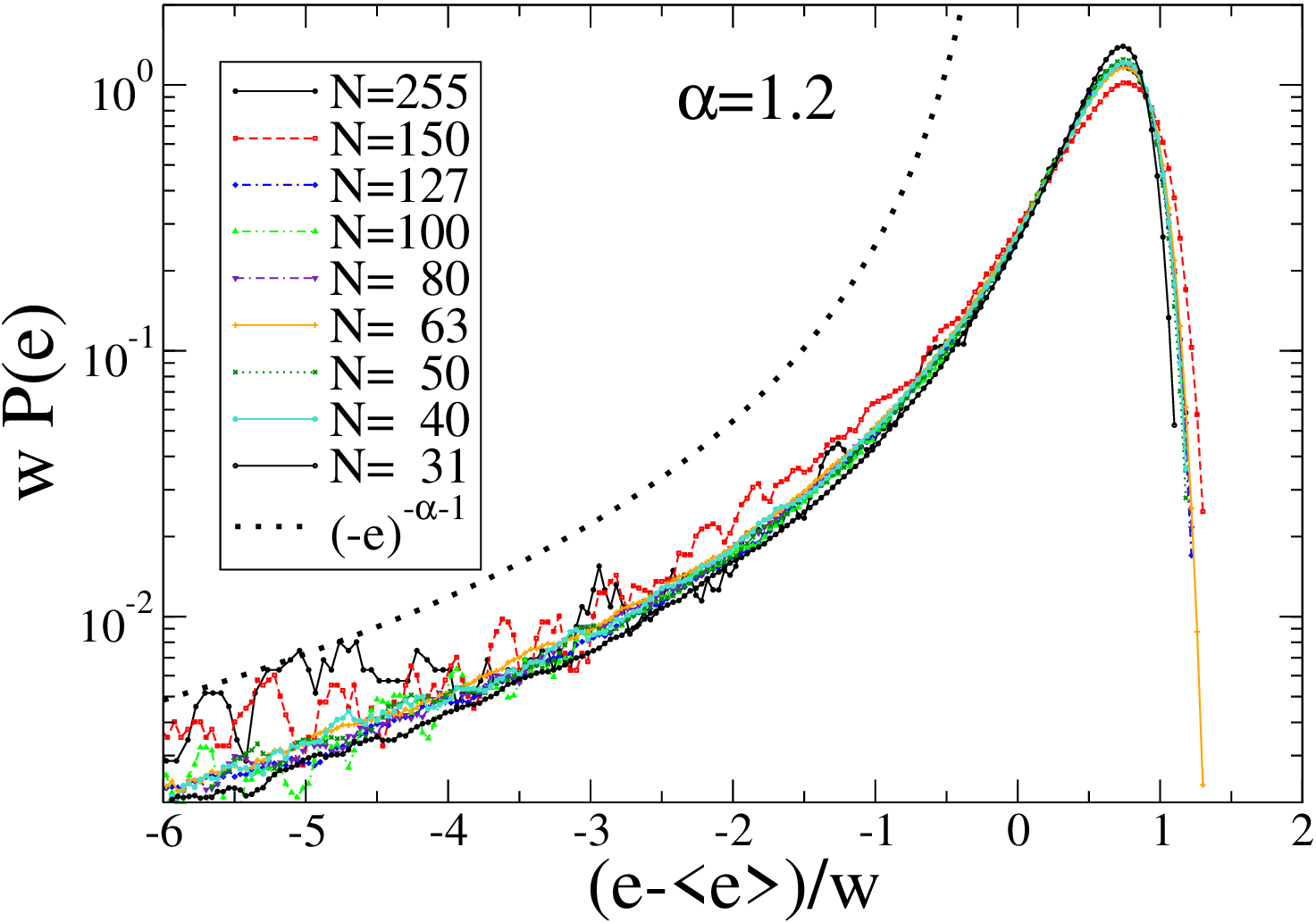}\includegraphics[scale=0.25]{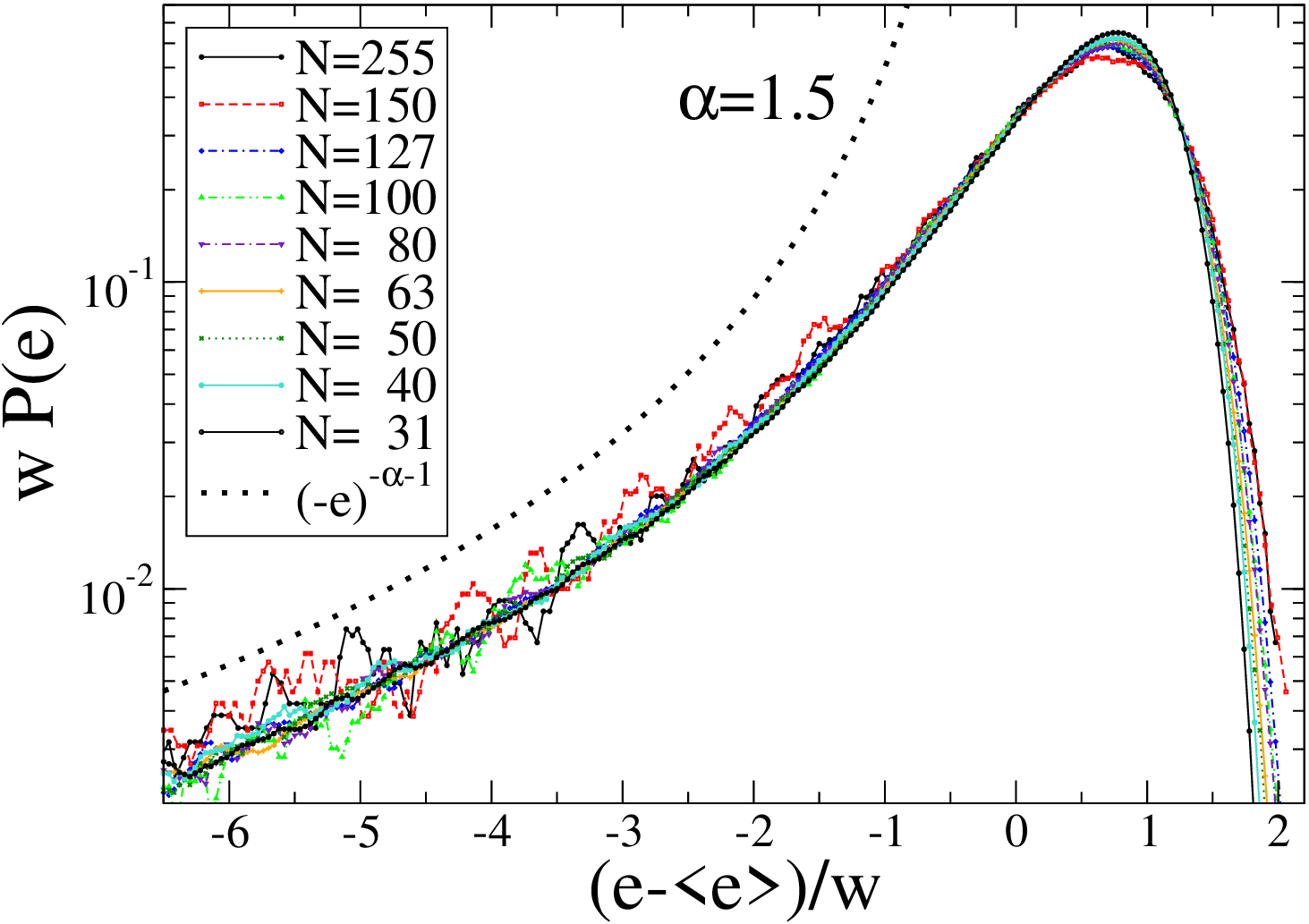}

\includegraphics[scale=0.25]{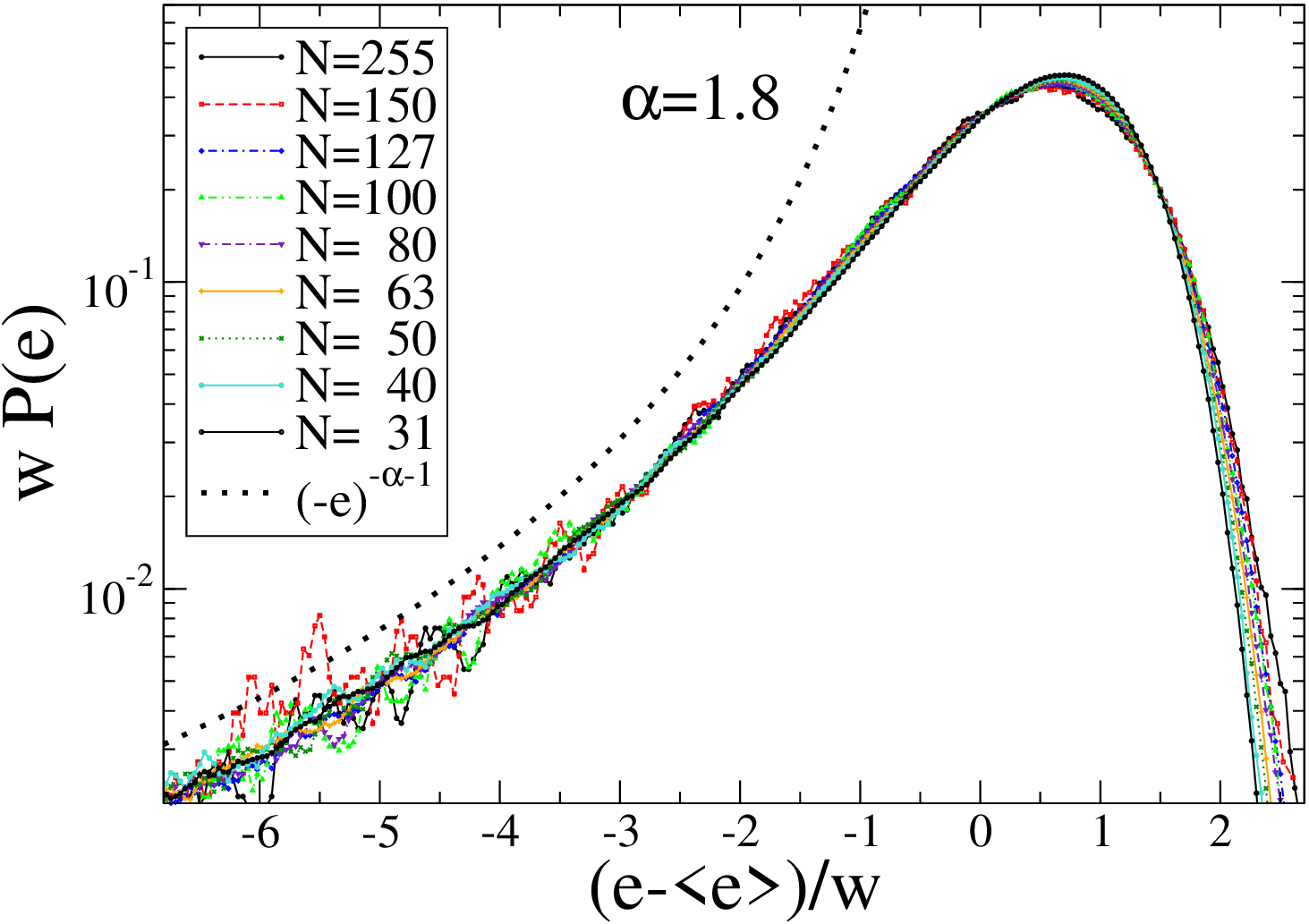}\includegraphics[scale=0.25]{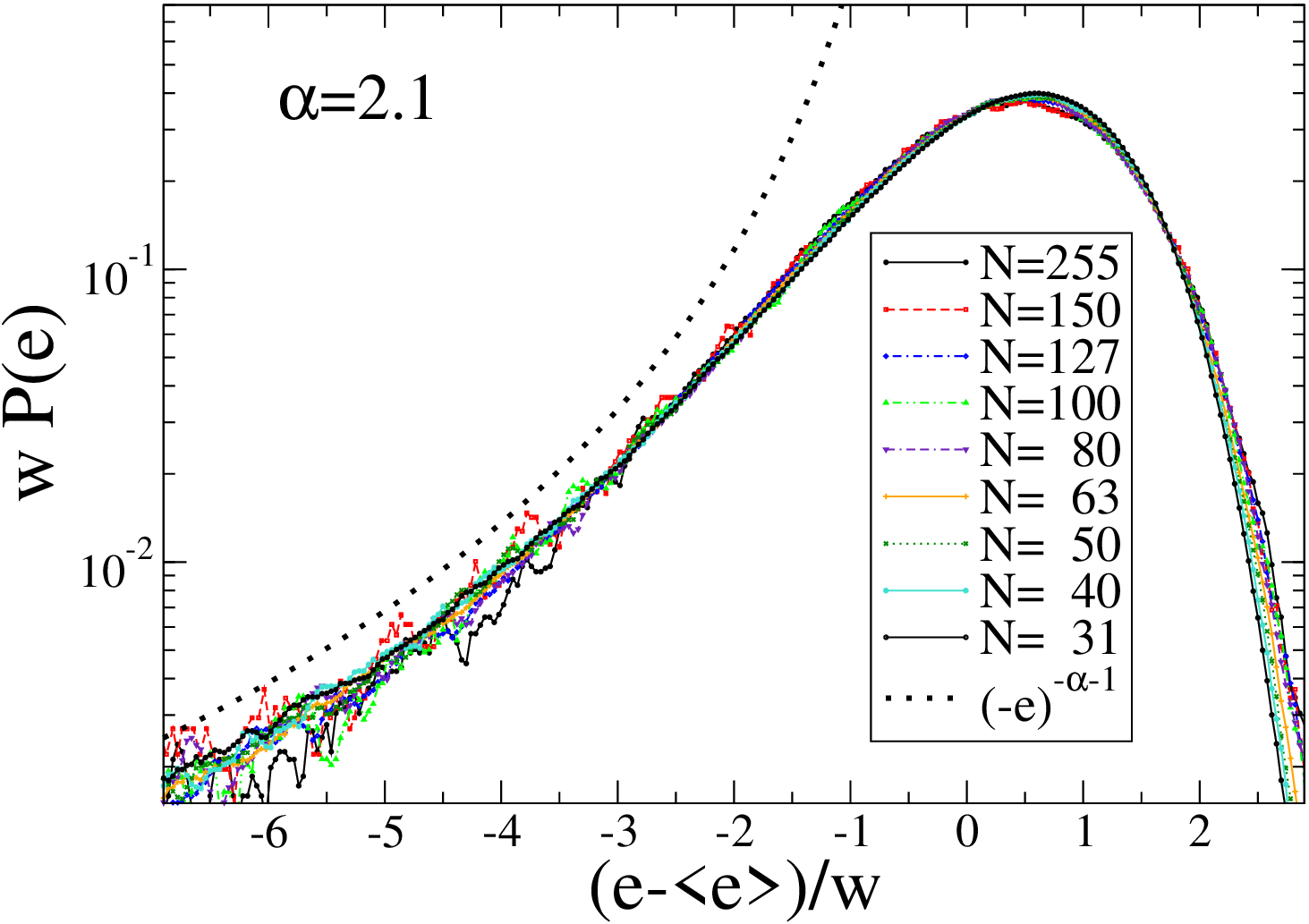}\includegraphics[scale=0.25]{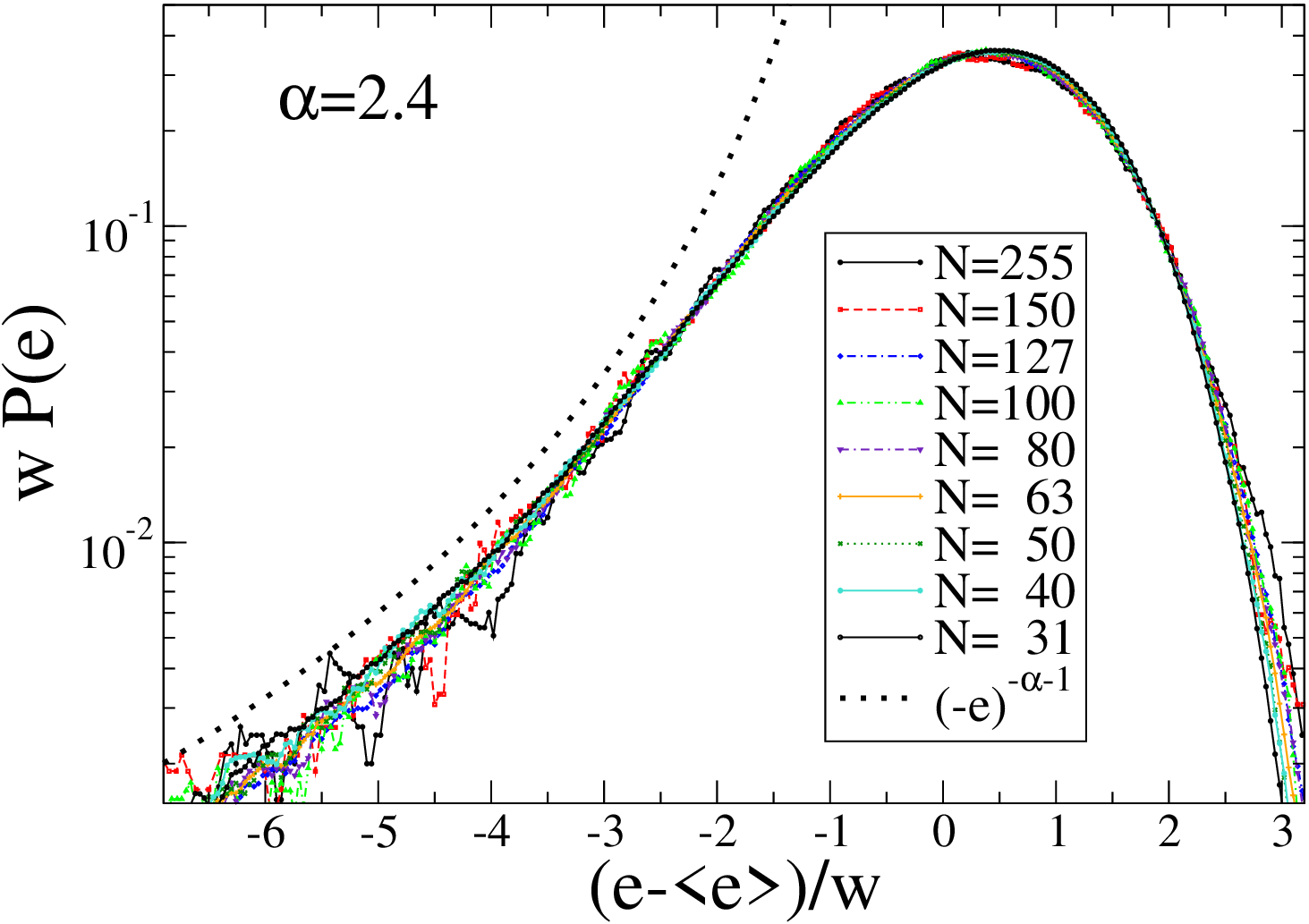}

\includegraphics[scale=0.25]{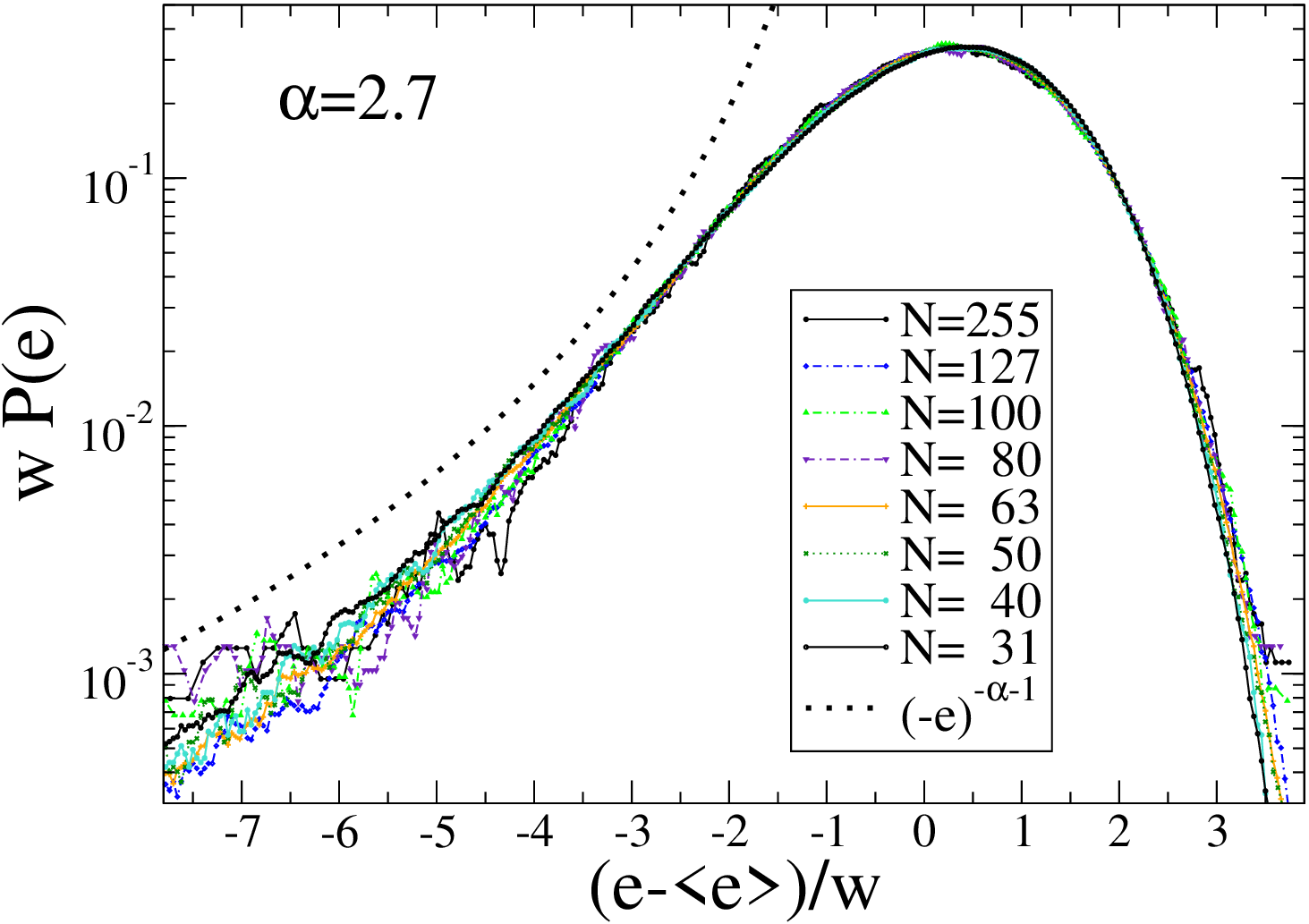}\includegraphics[scale=0.25]{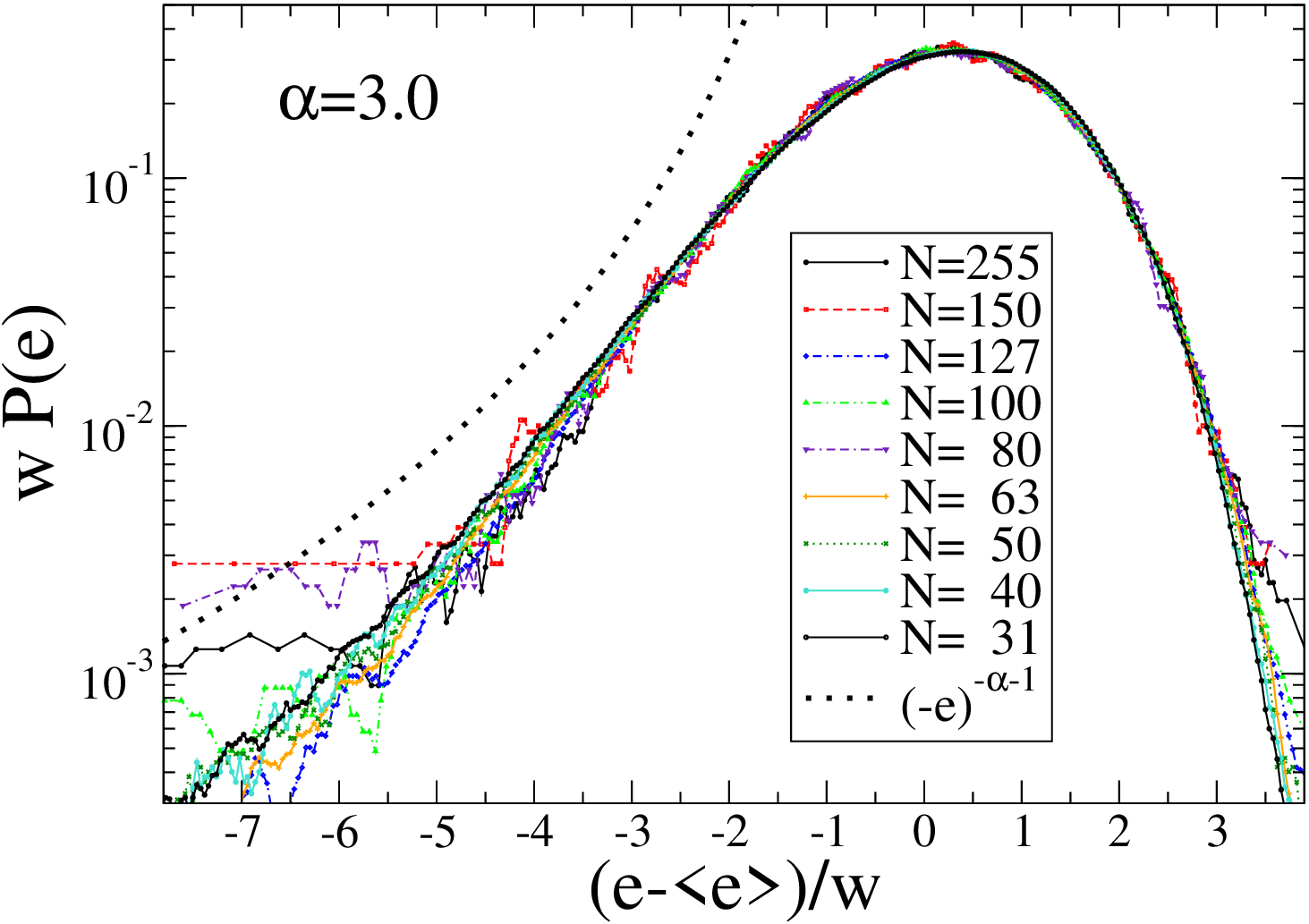}\includegraphics[scale=0.25]{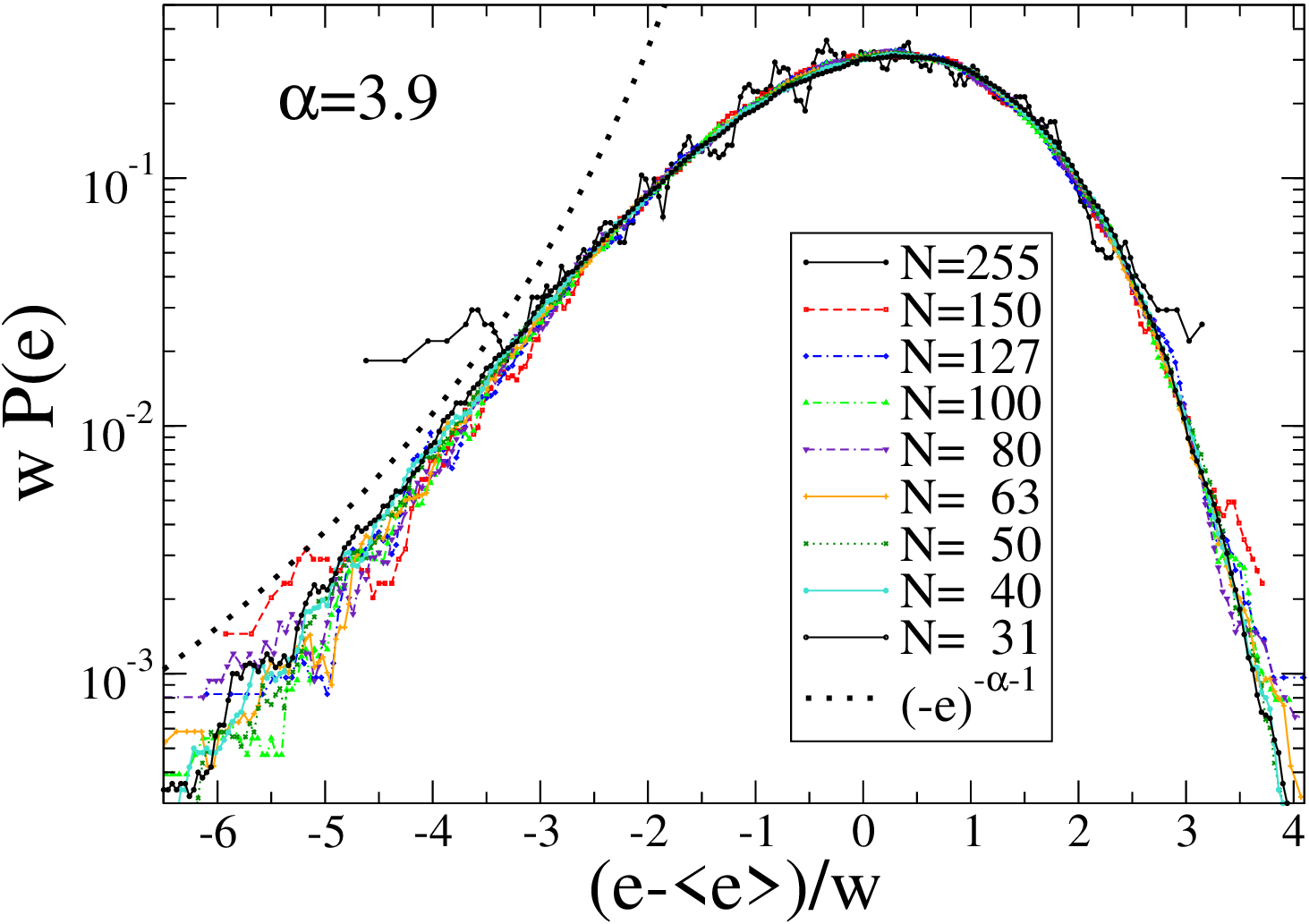}
\caption{\label{fig:PDF}Plot of the ground state energy PDF obtained
  with EO. We plot the data for all values of $N$ in each panel,
  whereas $\alpha$ varies between panels from 0.9 to 3.9. The PDFs
  appear to reach an asymptotic form quickly with small finite-size
  effects, which are most pronounced for smaller $\alpha$ (and, in
  fact, are divergent for $\alpha<1$). For $\alpha<2$, the tails for
  $e\rightarrow-\infty$ are a power-law (dotted lines) with the same
  exponent as the input bond-distribution $P(J)$ in
  Eq.~(\ref{eq:powerlaw}), since those ground state energies are
  dominated by a few large bonds that must be satisfied. For
  $\alpha>2$, the random sum of many small bonds dominates the energy,
  leading to increasingly exponential tails and the PDF becomes
  similar to that found in
  SK~\citep{Andreanov04,EOSK,Boettcher05e,Palassini08}.  The scaling
  of the width $w$ with system size is discussed in Fig.~\ref{fig:rhoplot}.}
\end{figure*}

\subsection{Ground-state energy fluctuations}

A deeper insight into the subtleties of ground state energies is
provided by their distribution over the ensemble of
instances. Typically, we plot the PDF as $\sigma P(e)$
vs. $(e-\left\langle e\right\rangle )/\sigma,$ where
$\sigma=\sqrt{\left\langle e^{2}\right\rangle -\left\langle
  e\right\rangle ^{2}}$ is the standard deviation of the
distribution. But for $\alpha<2,$ we find that the PDF has a broad
tail for energies \emph{below} the mean, which behaves as
$P(e)\sim(-e)^{-\alpha-1}$ for $e\rightarrow-\infty.$ Hence,
$\left\langle e^{2}\right\rangle $ and $\sigma$ do not exist.
Instead, we define a width via the absolute moment (see
Ref.~\citep{press:95}) $w=\left\langle \left|e-\left\langle
e\right\rangle \right|\right\rangle $ to characterize the distribution
and plot $wP(e)$ vs. $(e-\left\langle e\right\rangle )/w$ for each
$\alpha$ in Figs.~\ref{fig:PDF}.

The power-law tails at low energies $e\rightarrow-\infty$ in Figs.
\ref{fig:PDF} for $\alpha<2$ are easy to explain: They are entirely
due to those rare, large bonds from deep within the bond distribution
in Eq.~(\ref{eq:powerlaw}), which almost always must be satisfied and
completely dominate the ground state energy when they occur
\citep{Newman94,Cieplak94}.  Accordingly, those tails of the PDF decay
with the same exponent, $1+\alpha,$ as Eq.~(\ref{eq:powerlaw}).

\begin{figure}
\includegraphics[scale=0.32]{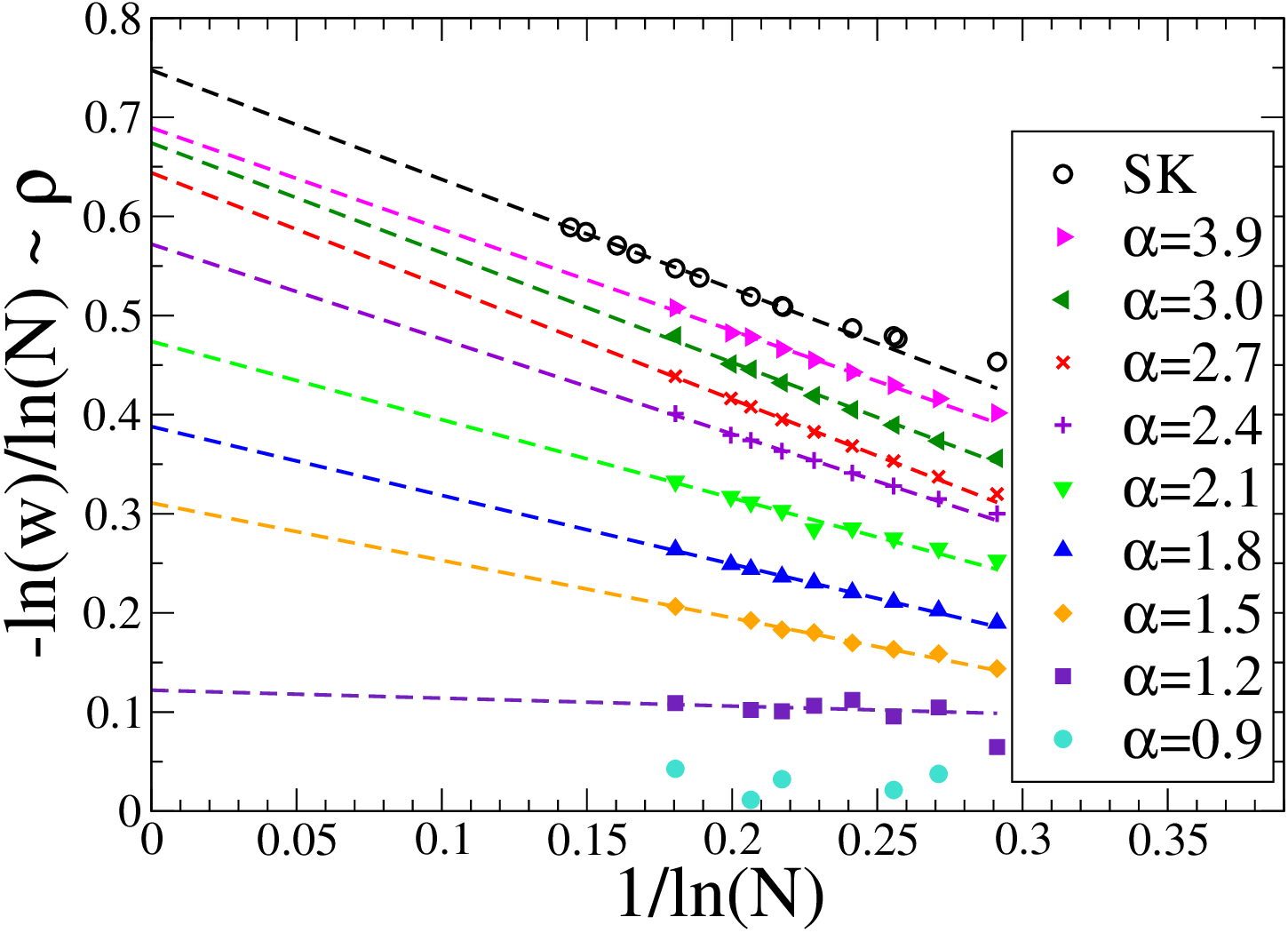}

\caption{\label{fig:rhoplot}Extrapolation plot of the width
  $w=\left\langle \left|e-\left\langle e\right\rangle
  \right|\right\rangle $ of the ground state energy distributions
  $P(e)$ in Figs.~\ref{fig:PDF} versus system size $N$ for each value
  of $\alpha$. Note that $\ln(w)/\ln(N)\to-\rho$ for
  $1/\ln(N)\to0$. As a reference, we have included the corresponding
  SK data for $w$ derived from the study in Ref.~\citep{Boettcher10b}.
  Theory~\citep{Parisi08} provides an upper bound of
  $\rho=\frac{5}{6}$ that should be approached for large $\alpha$. On
  the other side, when $\alpha$ drops below 1 the width diverges (in
  the limit of infinitely many instances), independent of system size
  $N$, leading to the conclusion $\rho_{\alpha\leq1}\equiv0$. For
  increasing $\alpha>1$, $\rho$ seems to increase rapidly to saturate
  soon at the SK limit, see  Fig.~\ref{fig:omegarho}.}
\end{figure}

The feature of $P(e)$ of greatest interest is the scaling of its
width. As noted, the standard deviation $\sigma$ does not exist for
$\alpha<2$, but we can instead for all $\alpha>1$ refer to the width
derived from the first absolute moment, $w=\left\langle
\left|e-\left\langle e\right\rangle \right|\right\rangle ,$ introduced
above. It is expected that, like $\sigma$, the width would decay with
system size as
\begin{eqnarray}
w & \sim & N^{-\rho}.
\label{eq:rho}
\end{eqnarray}
As an example of the significance of this exponent we mention that
according to Ref.~\citep{Aspelmeier03}, $\rho$ is related to the
exponent $y$ (or $\theta$) for domain-wall excitations in the
large-dimensional limit $d\to\infty$ of the Edwards-Anderson
model~\citep{Edwards75} via
\begin{equation}
y/d=1-\rho,
\label{eq:y_d}
\end{equation}
which has led to much consideration recently
\citep{Boettcher05d,Aspelmeier07,Parisi08,Aspelmeier08c,aspelmeier08b,Rizzo09b,Parisi09,Boettcher10b}.
In Fig.~\ref{fig:rhoplot}, we have plotted the values of $w$ obtained
in preparing Fig.~\ref{fig:PDF} as a function of $N$ for each value of
$\alpha$. In reference to Eq.~(\ref{eq:rho}), we specifically use its
extrapolated version and plot $-\ln(w)/\ln(N)$ as a function of
$1/\ln(N)$, which should extrapolate linearly to the thermodynamic
value of $\rho$ at the intercept $1/\ln(N)\to0$. It appears that
$\rho$ increases from zero at $\alpha\leq1$ to saturate at the SK
value. Although large uncertainties in the precise value of $\rho$ for
each individual $\alpha$ should be $ $expected (and possibly
non-analyticities or logarithmic corrections), the general trend in
$\rho$ appears to be reliable. The SK data provides justification for
choosing $w$ as the width instead of the more conventional deviation
$\sigma$: The extrapolation of $w$ in Fig.~\ref{fig:rhoplot} leads to
identical results for $\sigma$ as in Refs.~\citep{EOSK,Boettcher10b}.

\begin{figure}
\includegraphics[scale=0.36]{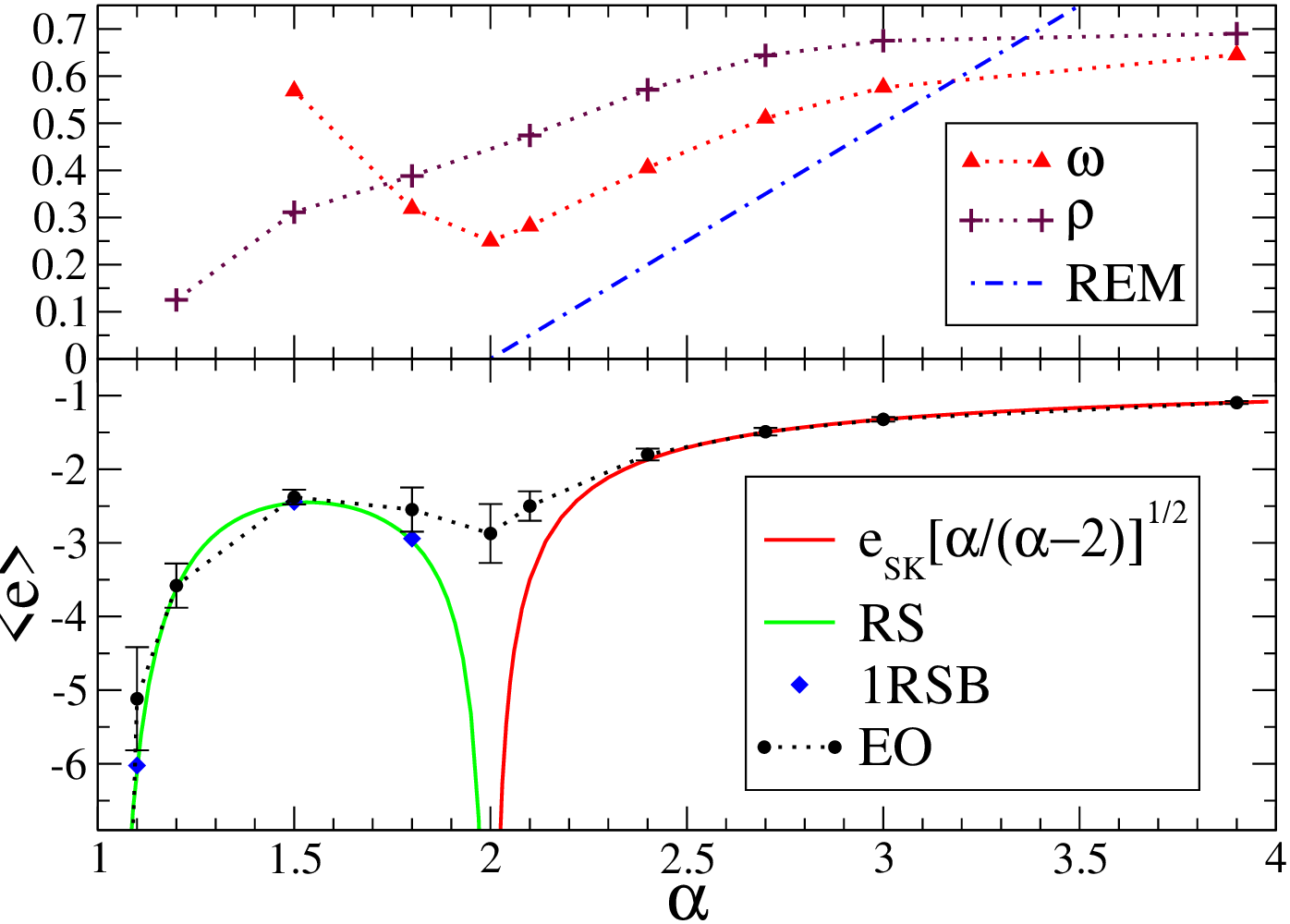}

\caption{\label{fig:omegarho}Top panel: Plot of the measured values of
  the exponents $\omega$ (from Figs.~\ref{fig:averageE}) and $\rho$
  (from Fig.~\ref{fig:rhoplot}) over the range of values for $\alpha$
  used in this study. (Lines are added just to guide the eye.) For
  $\alpha>1$, $\omega$ initially drops towards a minimum near
  $\alpha=2$ to rise again and approach its SK-value of 2/3. In turn,
  $\rho$ rises monotonically from about zero at $\alpha=1$ to its
  SK-value, which is which varies only little for $\alpha\geq3$. The
  dash-dotted line corresponds to the naive REM calculation, see
  Eq.~(\ref{eq:omega2<a<4}). Bottom panel: Plot of the ground state
  energy densities in the scaling of
  Eq.~(\ref{eq:energydensity}). Black circles denote the extrapolated
  EO values found in Fig.~\ref{fig:averageE} (connected by a dotted
  line to guide the eye). The line for $1<\alpha<2$ corresponds to the
  RS- and the blue diamonds to the 1RSB-calculation from
  Ref.~\citep{Janzen10a,Janzen10b}, the line for $\alpha>2$ provides
  the exact SK-energy $e_{SK}=-0.76321\ldots$, appropriately
  rescaled. The energies found with EO deviate significantly from the
  theoretical values near $\alpha=2$, where finite-size corrections
  are also the strongest (or, $\omega$ is minimal, see top).}

\end{figure}

\subsection{Exponents as a function of $\alpha$ for $2<\alpha<4$\label{sub:Exponents-as-a}}

While $\rho$ appears to rise rapidly but monotonously from zero at
$\alpha=1$ to approach the SK-limit, $\omega$ instead has a distinct
minimum of about $\omega\approx0.25$ near $\alpha=2$. For larger
$\alpha$, it approaches the presumed SK-value of $\omega=2/3$. In
turn, in the limit for $\alpha\to1^{-}$, $\omega$ may revert to
its SK value, although a simple volume-size correction to the energy
with $\omega=1$ or even exponentially small size corrections with
$\omega\to\infty$ appear conceivable. Although both exponents, $\omega$
and $\rho$, approach the corresponding SK-value convincingly for
larger values of $\alpha$, that limit is attained in a manner that
requires some explanation. Even for values of $\alpha>2$, where a
second moment in $P(J)$ already exists, both exponents still deviate
significantly from their SK-values. The smallest value beyond which
one might argue that the SK-limit has been saturated would be $\alpha=3$,
but it may even be higher. Our data would indicate a steady approach
to that limit but its system-size limitations certainly could not
exclude a singular {}``bend'' at $\alpha=3$, say.

We argue that the origin of these anomalous exponents for $\alpha>2$
can be tied to higher-order differences between the moments of a Gaussian
and a Levy distribution. In particular, for $\alpha\leq4$, the 4th
moment of the Levy distribution remains divergent. As both exponents
refer to finite-size effects, i.e., they do not per-se relate to properties
of the SK (that \emph{are} universal for $2<\alpha<4$) in the thermodynamic
limit, the sensitivity of these exponents to such differences is not
surprising. In the Appendix, we have done a speculative calculation
for the finite-size scaling corrections in the REM model~\citep{derrida:80,derrida:81}
with Levy bonds. The key result, Eq.~(\ref{eq:omega2<a<4}), clearly
demonstrates that at this level there are anomalous scaling exponents
already for $2<\alpha<4$, well in qualitative accord with our numerical
findings in Fig.~\ref{fig:omegarho}, where we indicated the REM result
by a dash-dotted line.

\begin{figure}
\includegraphics[scale=0.33]{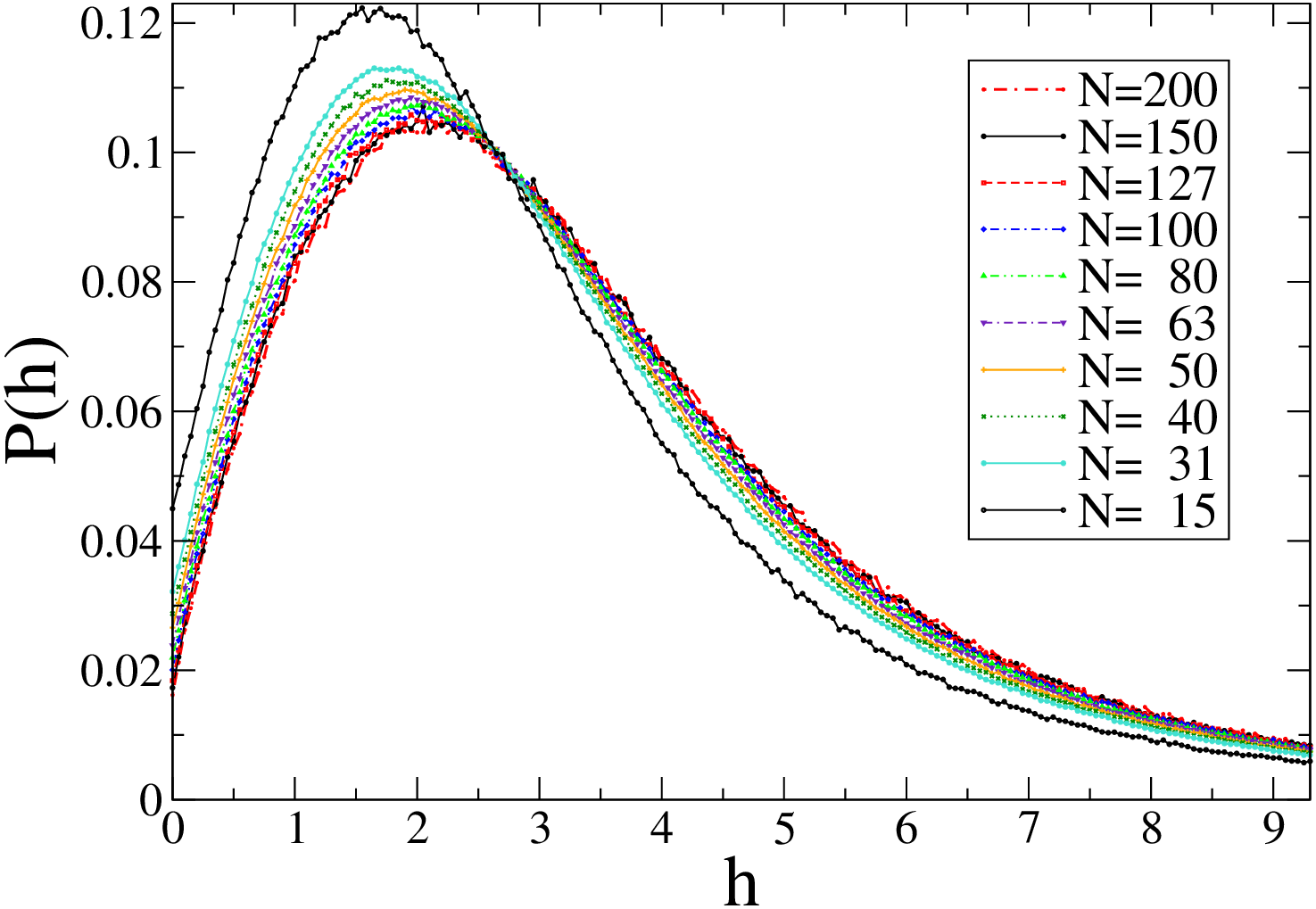}

\caption{\label{fig:Ph}Plot of the distribution of local fields $P(h)$ for
$h\geq0$ only for a range of system sizes $N$. (The function is
symmetric, and it is normalized to 1/2 for the positive part shown
here.) The distribution is nearly linear for $h\to0$, although higher
order terms become more noticeable for increasing $h$ than is the
case for SK~\citep{Boettcher07b}. For larger values of $h$, we observe
a much broader tail than for SK, yet, that tail still represents an
exponential fall-off. From this plot it is not clear whether $P(h)$
at $h=0$ remains finite for $N\to\infty$, therefore, we study the
behavior of $P(0)$ more closely in Fig.~\ref{fig:P0}.}

\end{figure}

\subsection{Local field distribution at $\alpha=3/2$ }

We now investigate also the distribution of local fields $P(h)$ in
the ground state configurations for $\alpha=3/2$. Instead of parsing
out the whole parameter space, we focus on a single value of $\alpha$
to optimize statistics. Specifically, $\alpha=3/2$ is chosen, since
it is ideally located sufficiently below the cross-over regime near
$\alpha=2$ , above which the central limit theorem comes into play,
and sufficiently above $\alpha=1$, where (numerical and theoretical)
pathologies arise due to the extreme breadth of the bond distribution.

In Fig.~\ref{fig:Ph} we plot the data obtained for $P(h)$ for $\alpha=3/2$
up to $N=200$. The function has similar characteristics to those
observed in other spin glasses~\citep{Boettcher07b}, with a near-Poissonian
shape, but with a much broader tail for larger $h$. Overall, at $N=200$
finite-size effects have largely diminished already, with the notable
exception near $h=0$, which in turn harbors the dynamically most
relevant information contained in $P(h)$. In the ground state configuration
(in which almost all variables have a positive local field) the number
of variables with near-vanishing local field characterize the stability
of the state, for which the scaling-behavior of $P(h)$ for $N\to\infty$
and $h\to0$ holds the key. Despite significant curvature for increasing
$h$, it seems clear from Fig.~\ref{fig:Ph} that the slope of $P(h)$
is nonetheless linear at the origin, as it is for SK.

\begin{figure}
\includegraphics[scale=0.32]{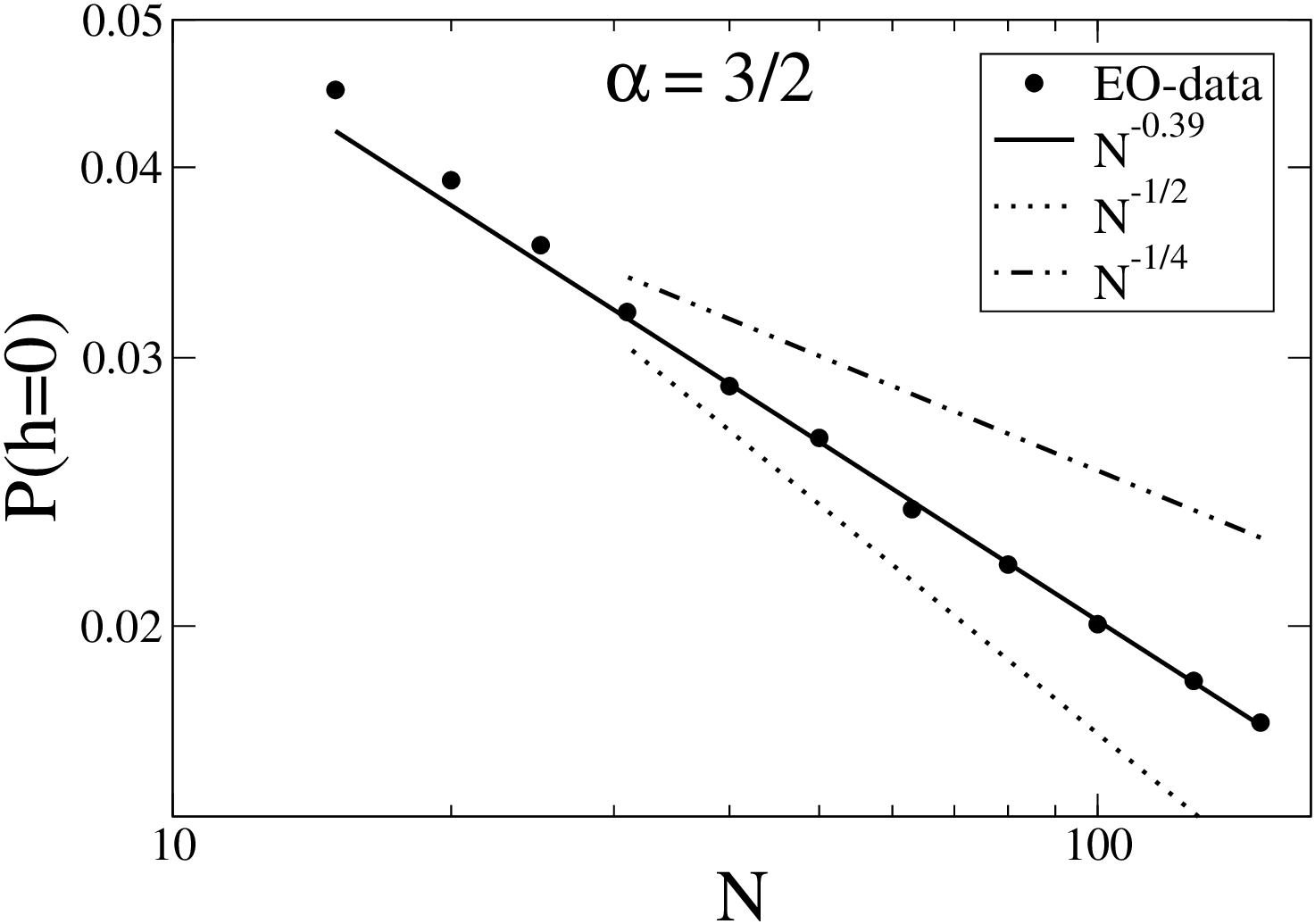}

\caption{\label{fig:P0}Plot of $P(0)$ as a function of system size $N$.
The data exhibits scaling over almost a decade, decaying roughly as
$N^{-0.39}$, displayed as the full line. For comparison, lines for
a $N^{-1/2}$ (dotted line) and $N^{-1/4}$-decay (dash-dotted line)
are also plotted.}

\end{figure}

To obtain a closer insight into the scaling of $P(0)$ itself, we
plot in Fig.~\ref{fig:P0} just the values at the origin of Fig.~\ref{fig:Ph}
as a function of system size $N$. It is very hard to get well-converged
data for larger system sizes, such as those available for SK~\citep{Boettcher07b},
but luckily the data exhibits already solid scaling starting with
$N\geq31$. The scaling, with an exponent of about $\approx0.4$,
indicates that $P(0)\to0$ in the thermodynamic limit but is definitely
slower decaying than for SK, which falls with $1/\sqrt{N}$~\citep{Boettcher07b}. 

\section{Acknowledgments\label{sec:Conclusions}}

Many thanks to J-P. Bouchaud for his suggestions that initiated this
project. I am greatly indebted to K. Janzen and A. Engel for many
clarifying discussion and for providing me with some of their data,
and to the Fulbright Kommission for supporting my stay at Oldenburg
University. This work has been supported also by grants 0312510 and
0812204 from the Division of Materials Research at the National Science
Foundation and by the Emory University Research Council.

\section{Appendix:\label{sec:Appendix:} Finite-Size Corrections in the Levy-REM}

We use here the assumption of the random energy model (REM)~\citep{derrida:80,derrida:81}
that the spectrum of available states are uncorrelated to obtain an
estimate for the finite-size scaling of the SK model with Levy bonds.
This is true for the $p$-spin generalization of SK in the limit of
$p\to\infty$ with Gaussian bonds, but this is clearly not obvious
for Levy bonds. We go even further and assume this to hold even for
$p=2$, the case considered for SK in this paper, in order to estimate
leading-order scaling and finite-size corrections to the ground state
energy density. A full treatment of the Levy-REM would likely be a
lengthy exercise but potentially rewarding as interesting classes
of the extreme-value distribution of ground-state energies and of
the local field distribution may result.

\subsection{REM with Gaussian bonds\label{sec:REM-with-Gaussian}}

To review the REM\citep{derrida:80,derrida:81,Gross84} calculation
for the Hamiltonian of the $p$-spin model,
\begin{eqnarray}
{\cal H}\left(\left\{ \sigma_{i}\right\} \right) & = &
\frac{1}{p!}\sum_{i_{1}=1,\ldots,i_{p}=1}^{N}J_{i_{1},\ldots,i_{p}}\sigma_{i_{1}}\times\ldots\times\sigma_{i_{p}},
\label{eq:pHamiltonian}
\end{eqnarray}
for Gaussian bonds, we set 
\begin{eqnarray}
P\left(J\right) & = & \sqrt{\frac{N^{p-1}}{\pi p!}}\exp\left\{
-\frac{N^{p-1}}{p!}J^{2}\right\} 
\label{eq:GaussianP}
\end{eqnarray}
and, following Ref.~\citep{F+H}, calculate the densities of energy-levels via
\begin{eqnarray}
&&{\cal D}\left(E\right)  =  \left\langle \delta\left(E-{\cal H}\left(\left\{ \sigma_{i}\right\} \right)\right)\right\rangle _{J},
\label{eq:PE}\\
 && =
  \int\prod_{i_{1},\ldots,i_{p}}^{N}\left[dJ_{i_{1},\ldots,i_{p}}P\left(J_{i_{1},\ldots,i_{p}}\right)\right]\nonumber\\
&&\qquad\delta\left(E-\frac{1}{p!}\sum_{i_{1},\ldots,i_{p}}^{N}J_{i_{1},\ldots,i_{p}}\sigma_{i_{1}}\ldots\sigma_{i_{p}}\right),\nonumber \\
 && = 
\int_{-\infty}^{\infty}\frac{dt}{2\pi}\,e^{iEt}\prod_{i_{1}=1,\ldots,i_{p}=1}^{N}\left[\int
  dJP\left(J\right)\exp\left\{\frac{iJt}{p!}\sigma_{i_{1}}\ldots\sigma_{i_{p}}\right\}\right],
\nonumber 
\end{eqnarray}
assuming in the first line that the energies are uncorrelated:
\begin{equation}
{\cal D}\left(E_{1},E_{2}\right)={\cal D}\left(E_{1}\right){\cal
  D}\left(E_{2}\right).
\label{eq:Dfactorization}
\end{equation}
The ensuing Gaussian integral in Eq.~(\ref{eq:PE}) only depends on
$\left(\sigma_{i_{1}}\ldots\sigma_{i_{p}}\right)^{2}\equiv1$,
thus,
\begin{equation}
{\cal D}\left(E\right) = \int_{-\infty}^{\infty}\frac{dt}{2\pi}e^{N\psi\left(t\right)}
\label{eq:psi_integral}
\end{equation}
with
\begin{equation}
\psi\left(t\right)  =  it\frac{E}{N}-\frac{t^{2}}{4}.
\end{equation}
Of course, no steepest-descent analysis is required here to solve
\emph{this} integral:
\begin{eqnarray}
{\cal D}\left(E\right) & = & \frac{e^{-\frac{E}{N}^{2}}}{\sqrt{N\pi}}.\label{eq:D-REM}\end{eqnarray}
Then, we get for the entropy
\begin{eqnarray}
S\left(E\right)  &=&  \ln\Omega\left(E\right) 
= \ln\left[2^{N}{\cal D}\left(E\right)\right]\nonumber\\
 &=&  N\left[\ln2-\left(\frac{E}{N}\right)^{2}+O\left(\frac{\ln N}{N}\right)\right].
\label{eq:REMentropyGauss}
\end{eqnarray}
As the entropy may not become negative, vanishing of the entropy (-density)
defines the ground-state energy density
\begin{eqnarray}
\frac{E_{0}}{N} & = & -\sqrt{\ln2}+O\left(\frac{\ln N}{N}\right).
\label{eq:REM-GS-Gauss}
\end{eqnarray}
Its numerical value, $-\sqrt{\ln2}=-0.832555$, is plausible when
compared to the Parisi energy, $e_{SK}=-0.763217\ldots$
\citep{Oppermann05,Oppermann07}.  The implicated finite-size
correction of this result, $\ln\left(N\right)/N$ or $\omega=1$, of
course, does not correspond to the SK result for $p=2$, presumed to
scale with $N^{-\frac{2}{3}}$. But since this result has become
entirely independent of $p\geq2$, one would hardly expect a better
agreement. After all, only for $p\to\infty$ are these energy levels
sufficiently uncorrelated, as in Eq.~(\ref{eq:Dfactorization}), to
justify this approach. The value of our calculation here lies not so
much in a precise prediction for $\omega$ but in a plausible trend in
the function $\omega(\alpha)$ and its transitions.

\subsection{REM with Levy Bonds\label{sec:REM-with-Levy}}

We can repeat the above calculation for the $p$-spin Hamiltonian
in Eq.~(\ref{eq:pHamiltonian}) with a bond distribution of the
Levy-type,
\begin{eqnarray}
P\left(J\right) & = &
\frac{\alpha}{2}B^{\alpha}\left|J\right|^{-1-\alpha}\theta\left(\left|J\right|-B\right),
\label{eq:Levy}
\end{eqnarray}
for some positive $B$, which will be discussed below.

Inserting this bond distribution into Eq.~(\ref{eq:PE}) results in
a density of states ${\cal D}\left(E\right)$ as in Eq.~(\ref{eq:psi_integral})
using
\begin{equation}
\psi\left(t\right) = \pm it\frac{E}{N}+N^{p-1}\ln\left[f_{\alpha}\left(\frac{tB}{p!}\right)\right]
\label{eq:psi-f}
\end{equation}
with
\begin{equation}
f_{\alpha}\left(x\right) =
\alpha\int_{1}^{\infty}d\xi\frac{\cos\left[x\xi\right]}{\xi^{1+\alpha}},
\end{equation}
which can be written in terms of special functions to facilitate
the ensuing analysis:
\begin{eqnarray}
f_{\alpha}\left(x\right) & = &
\,{}_{1}F_{2}\left(-\frac{\alpha}{2};\frac{1}{2},1-\frac{\alpha}{2};-\frac{x^{2}}{4}\right)\nonumber\\
&&\quad-\cos\left(\frac{\pi}{2}\alpha\right)\Gamma\left(1-\alpha\right)x^{\alpha}.
\label{eq:f-hyper}
\end{eqnarray}
For even $\alpha=2,4,\ldots$, both terms in Eq.~(\ref{eq:f-hyper})
develop identical singularities for small $x$ in a way that the resulting
expression always remains finite:
\begin{eqnarray}
f_{2}\left(x\right) & = & \cos\left(x\right)-x\sin\left(x\right)+x^{2}{\rm Ci}\left(x\right),\label{eq:f24}\\
f_{4}\left(x\right) & = &
\left(1-\frac{x^{2}}{6}\right)\cos\left(x\right)\nonumber\\
&&\quad-\frac{x}{6}\left(2-x^{2}\right)\sin\left(x\right)-\frac{x^{4}}{6}{\rm
  Ci}\left(x\right).\nonumber 
\end{eqnarray}

From Eq.~(\ref{eq:psi-f}) it is clear that for the saddle point in
the thermodynamic limit $N\to\infty$, $B$ will have to be chosen
to scale with the system size as to make $\psi\left(t\right)$ stationary
in that limit and to render the energy density $E/N$ intensive, i.
e., it has to serve to compensate the explicit factor of $N^{p-1}$.
We expect that for all values of $\alpha$ the saddle point is located
at some (possibly complex) finite value of $t_{0}$ while $B$ becomes
small in some fashion for $N\to\infty$. In that case, we are merely
looking for the series expansion of $\ln\left[f_{\alpha}\left(x\right)\right]$
for small values of $x$:
\begin{eqnarray}
-\ln f_{\alpha}\left(x\right) & \sim& 
\cos\left(\frac{\pi}{2}\alpha\right)\Gamma\left(1-\alpha\right)x^{\alpha}+\frac{\alpha}{2\left(\alpha-2\right)}x^{2}\nonumber\\
&&\quad+\frac{\alpha\left(\alpha^{2}-4\alpha-2\right)}{12\left(\alpha-4\right)\left(\alpha-2\right)^{2}}x^{4}+\ldots,
\label{eq:lnfalpha}
\end{eqnarray}
as long as $\alpha>2$; for $\alpha<2$ there are higher powers of
the $x^{\alpha}$-term in Eq.~(\ref{eq:f-hyper}) from the expansion
of the logarithm to take into account. There are exceptional cases
for $\alpha=2,4,\ldots$:
\begin{eqnarray}
\ln f_{2}\left(x\right) & \sim & x^{2}\left[\left(\ln
  x+\gamma\right)-\frac{3}{2}\right]\nonumber\\
&&-\frac{x^{4}}{6}\left[3\left(\ln
  x+\gamma\right)^{2}-9\left(\ln x+\gamma\right)+7\right]+\ldots,
\nonumber \\
\ln f_{4}\left(x\right) & \sim &
-x^{2}-\frac{x^{4}}{72}\left[12\left(\ln  x+\gamma\right)+11\right]+\ldots.
\label{eq:lnf24}
\end{eqnarray}

\paragraph{Case $\alpha>4$:\label{sub:a>4}}

For $\alpha>4$, the irregular term in Eq.~(\ref{eq:lnfalpha}) becomes
irrelevant for the determination of ground-state energy density and
its corrections. We find for $\psi\left(t\right)$ in Eq.~(\ref{eq:psi-f})
to leading orders,
\begin{eqnarray}
\psi\left(t\right) & \sim &
it\frac{E}{N}-\frac{\alpha}{2\left(\alpha-2\right)}N^{p-1}\left(\frac{tB}{p!}\right)^{2}\nonumber\\
&&-\frac{\alpha\left(\alpha^{2}-4\alpha-2\right)}{12\left(\alpha-4\right)\left(\alpha-2\right)^{2}}N^{p-1}\left(\frac{tB}{p!}\right)^{4}+\ldots,
\nonumber\\
 & \sim & it\frac{E}{N}-\frac{t}{4}^{2}+C\frac{t^{4}}{N^{p-1}}+\ldots,
\label{eq:psi>4}
\end{eqnarray}
for some constant $C$ and with the choice of 
\begin{eqnarray}
B & = & \sqrt{\frac{p!\left(\alpha-2\right)}{2\alpha}}\,N^{-\frac{p-1}{2}}.
\label{eq:B-SK}
\end{eqnarray}
We can write the density of states in Eq.~(\ref{eq:psi_integral}) as
\begin{eqnarray}
{\cal D}\left(E\right) & \sim &
\int_{-\infty}^{\infty}\frac{dt}{2\pi}\,e^{N\left(it\frac{E}{N}-\frac{t}{4}^{2}\right)+C\frac{t^{4}}{N^{p-2}}+\ldots}.
\label{eq:psi-integral>4}
\end{eqnarray}
Focusing only on a small $\epsilon$-neighborhood in $t$ near the
saddle-point $t_{0}=2i\frac{E}{N}$ (with $1/\sqrt{N}\ll\epsilon\ll1$),
followed by the shift $t=t_{0}+\frac{u}{\sqrt{N}}$ yields~\citep{BO}
\begin{eqnarray}
{\cal D}\left(E\right) 
 & \sim &
\frac{e^{-N\left[\left(\frac{E}{N}\right)^{2}+O\left(\frac{1}{N^{p-1}}\right)\right]}}{\sqrt{\pi
    N}}
\label{eq:DE>4}\\
&&\quad\int_{-\epsilon\sqrt{N}\to-\infty}^{\epsilon\sqrt{N}\to\infty}\frac{du}{\sqrt{4\pi}}\,\left[1+O\left(\frac{u^{2}}{N^{p-1}}\right)\right]e^{-\frac{u^{2}}{4}},\nonumber\\
 & \sim & \exp\left\{
-N\left[\left(\frac{E}{N}\right)^{2}+O\left(\frac{\ln
    N}{N}\right)+O\left(\frac{1}{N^{p-1}}\right)\right]\right\} .
\nonumber
\end{eqnarray}
Hence, for $\alpha>4$ there is neither a change in scaling for the
leading-order calculation for the ground state energy density nor
for its finite size correction from those obtained for the REM with
Gaussian bonds in Sec.~\ref{sec:REM-with-Gaussian}, as for $p=2$
(or larger) any corrections arising from the broader tails remain
sub-dominant. Furthermore, the pre-factor in the choice of $B$ in
Eq.~(\ref{eq:B-SK}) ensures that the second moment of the bond distribution
remains unity, $\left\langle J^{2}\right\rangle =1$, for all $\alpha>2$,
such that the ground-state energy is also numerically identical to
that of the Gaussian case.

\paragraph{Case $2<\alpha<4$:\label{sub:Case2<a<4}}

In this case, the irregular term in Eq.~(\ref{eq:lnfalpha}) now becomes
relevant for the determination of the finite-size correction for the
ground-state energy density and its corrections. With the same choice
of $B$ as in Eq.~(\ref{eq:B-SK}), we then find for $\psi\left(t\right)$
in Eq.~(\ref{eq:psi-f}) to leading orders,
\begin{eqnarray}
\psi\left(t\right) & \sim &
it\frac{E}{N}-\frac{\alpha}{2\left(\alpha-2\right)}N^{p-1}\left(\frac{tB}{p!}\right)^{2}\nonumber\\
 &&~~+\cos\left(\frac{\pi}{2}\alpha\right)\Gamma\left(1-\alpha\right)N^{p-1}\left(\frac{tB}{p!}\right)^{\alpha}+\ldots,\nonumber \\
 & \sim &
it\frac{E}{N}-\frac{t}{4}^{2}+C\frac{t^{\alpha}}{N^{\left(\frac{\alpha}{2}-1\right)\left(p-1\right)}}+\ldots.
\label{eq:psi2<a<4}
\end{eqnarray}
Following exactly along the lines of Eqs.~(\ref{eq:psi-integral>4}-\ref{eq:DE>4}),
we obtain
\begin{equation}
-\frac{\ln{\cal D}\left(E\right)}{N}  \sim 
\left(\frac{E}{N}\right)^{2}+O\left(\frac{\ln
  N}{N}\right)+O\left(\frac{1}{N^{\left(\frac{\alpha}{2}-1\right)\left(p-1\right)}}\right),
\nonumber
\end{equation}
where now the correction in $\psi(t)$ accounts for the next-to-leading
term, dominating the factor originating from the Gaussian saddle-point
integral for some ($p>2$) or all ($p=2$) of this regime. We surmise
that to leading order we retain the ground-state energy from the Gaussian
case here (as the second moment of the bond distribution still exists!),
but already in this regime of $\alpha$ we find a non-universal effect
in terms of the finite-size corrections~\citep{Boettcher10a}. For
instance, ignoring the fact that statistical independence of the energy
levels only holds for $p\to\infty$, we boldly set $p=2$ to extract
an approximation for the finite-size scaling exponent, \begin{eqnarray}
\omega(\alpha) & = & \frac{\alpha}{2}-1\qquad\left(2<\alpha<4\right).\label{eq:omega2<a<4}\end{eqnarray}
While not in great quantitative agreement with the numerical results
for $p=2$, this analysis does indeed capture the essence of the numerical
results to some qualitative satisfaction.

\paragraph{Case $\alpha=2$:\label{sub:Case=00003D2}}

It seems clear that in this case not only the corrections but also
the ground state itself will pick up some messy form of log-scaling.
To this end, it suffices to determine the form of $B$ that leaves
the saddle-point stationary for $N\to\infty$. From Eqs.~(\ref{eq:psi-f})
and expanding~(\ref{eq:lnf24}) to sufficient order, we get\begin{eqnarray}
\psi\left(t\right) & \sim & it\frac{E}{N}+N^{p-1}\left(\frac{tB}{p!}\right)^{2}\ln\left(\frac{tB}{p!}\right)+\ldots.\label{eq:psi=00003D2}\end{eqnarray}
Even for general complex $t$, the saddle-point is still determined
via $\psi'\left(t_{0}\right)=0$. The obtained saddle-point at \begin{eqnarray}
t_{0} & \sim & i\left(\frac{E}{N}\right)\frac{\left(p!\right)^{2}}{2N^{p-1}B^{2}\ln\left(N^{p-1}B\right)}\label{eq:t0=00003D2}\end{eqnarray}
becomes stationary for the choice of \begin{eqnarray}
B & = & \frac{1}{N^{\frac{p-1}{2}}\sqrt{\ln N^{\frac{p-1}{2}}}}.\label{eq:B=00003D2}\end{eqnarray}
Thus, we have acquired an unusual logarithmic correction, already
for the leading behavior of the ground state energy density. Inserting
$t_{0}$ in Eq.~(\ref{eq:t0=00003D2}) into $\psi\left(t\right)$
in Eq.~(\ref{eq:psi=00003D2}) yields\begin{eqnarray}
\psi\left(t_{0}\right) & = & -\frac{\left(p!\right)^{2}}{4}\left(\frac{E}{N}\right)^{2}\left[1+O\left(\frac{\ln\sqrt{\ln N^{\frac{p-1}{2}}}}{\ln N^{\frac{p-1}{2}}}\right)\right],\label{eq:psit0=00003D2}\end{eqnarray}
assuming that any other contribution from the actual saddle-point
integral only results in much smaller corrections than the indicated
ones, arising from the corrections in the motion of the saddle-point
itself. At $p=2$, we would therefore predict for the ground state
energy density, \begin{eqnarray}
\frac{E_{0}}{N} & = & -\sqrt{\ln2}+O\left(\frac{\ln\sqrt{\ln\sqrt{N}}}{\ln\sqrt{N}}\right),\label{eq:REM-GS=00003D2}\end{eqnarray}
with the same thermodynamic limit as before, but much smaller corrections.
Of course, such a scaling is nearly impossible to verify numerically.

\paragraph{Case $1<\alpha<2$: \label{sub:Case1<2}}

Now, the saddle-point analysis changes somewhat, with the integration
contour being rotated into the complex plane, requiring a more sophisticated
steepest-descent approach~\citep{BO}. In our naive approach taken
here, we find that reasonable solutions for $E_{0}$ persist for $\alpha>\frac{3}{2}$,
at which point the saddle point rotates across the branch-cut on the
negative real-axis in the complex-$t$ plane. We leave this calculation
as an exercise for the reader.

\bibliographystyle{apsrev}
\bibliography{/Users/stb/Boettcher}

\end{document}